\documentclass[fleqn,usenatbib,useAMS]{mnras}

\usepackage{graphicx}    
    \usepackage{amsmath}    
    \usepackage{amssymb}  

\usepackage{newtxtext,newtxmath}

\newcommand{\Msun}{\mbox{$M_{\odot}$}}

\newcommand{\beq}{\begin{equation}}
\newcommand{\eeq}{\end{equation}}
\newcommand{\beqa}{\begin{eqnarray}}
\newcommand{\eeqa}{\end{eqnarray}}
\newcommand{\bite}{\begin{itemize}}
\newcommand{\eite}{\end{itemize}}

\newcommand{\comment}[1]{}

\title[The molecular disc bumped by the jet]{Digging process in NGC 6951: the molecular disc bumped by the jet}
\author[D. May et al.]{
D. May $^{1}$\thanks{E-mail: dmay@usp.br}, 
J.E. Steiner$^{1}$, T.V. Ricci$^{1}$, R.B. Menezes$^{1}$, I.S. Andrade$^{1}$
\\
$^{1}$Institute of Astronomy, Geophysics and Atmospheric Sciences, University of S\~ao Paulo\\ Rua do Mat\~ao 1226, Cidade Universit\'aria, S\~ao Paulo, SP CEP 05508-090, Brazil}
\begin{document}

\date{Draft for internal use only}

\pagerange{\pageref{firstpage}--\pageref{lastpage}} \pubyear{2014}

\maketitle

\label{firstpage}

\begin{abstract}
We present a study of the central 200 pc of the galaxy NGC 6951, SAB(rs)bc, an active twin of the Milky Way, at a distance of 24 Mpc. Its nucleus has been observed in the optical with the GMOS-IFU, showing an outflow, and with the HST/ACS, revealing two extended structures with similar orientation, suggesting the presence of a collimating and/or obscuring structure. In order to ascertain this hypothesis, adaptive optics assisted NIR integral field spectroscopic observations were obtained with the NIFS spectrograph in the Gemini North telescope. We detected a compact structure of H$_2$ molecular gas, interpreted as a nearly edge-on disc with diameter of $\sim$47 pc, PA=124\textdegree~and velocity range from -40 to +40 km s$^{-1}$. This disc is misaligned by 32\textdegree~with respect to the radio jet and the ionization cones seen in the optical. There are two regions of turbulent gas, with position angles similar to the jet/cones, seen both in molecular and ionized phases; these regions are connected to the edges of the molecular disc and coincide with a high ratio of [N II]/H$\alpha$=5, suggesting that these regions are shock excited, partially ionized or both. We explain these structures as a consequence of a ``digging process'' that the jet inflicts on the disc, ejecting the molecular gas towards the ionization cones. The dynamical mass within 17 pc is estimated as $6.3\times10^{6}$\Msun. This is an interesting case of an object presenting evidence of a connected feeding-feedback structure.
\end{abstract}


\section{Introduction}
\label{sec:intro}

Active Galactic Nuclei (AGNs) comprise some typical components: a supermassive central black hole (SMBH), with a range of mass between $10^{6-10}$ \Msun; an accretion disc; a dusty and thick torus, responsible for the obscuration in some AGNs, with the internal radius determined by the sublimation radius of the dust; and, not always observable, a radio jet. This jet, supposedly, is launched from the internal parts of the disc, where the highest energy densities are located. The orientation of the jet is given by the structure of the inner accretion disc \citep{McKinney14}, which, in turn, may not be aligned with the outer gas disc, the source of fuel for the AGN \citep{Pringle03}.

AGNs can be classified as type 1, where the permitted lines are significantly broader than the forbidden lines, or type 2, where these lines present similar widths. The difference between types 1 and 2 is attributed to the geometric orientation of a dusty torus with respect to the line of sight (LOS) \citep[see the Unified Model by][]{Antonucci93}. According to this model, when the torus is seen edge-on, the Broad Line Region (BLR) is obscured and only the Narrow Line Region (NLR) is visible. However, intrinsic differences have been found between the two types, such as the absence of broad emission lines in the spectra of polarized light of Seyfert 2 galaxies \citep{Gu02}. Galaxies with Seyfert nuclei, of higher luminosity, present high ionization spectra, while Low Ionization Nuclear Emission-Line Regions (LINERS) \citep{Heckman80} are common among low luminosity AGNs (LLAGNs) \citep{Ho08}.

There is little doubt regarding the main points of the Unified Model scheme, but the exact nature of the torus is still a matter of discussion. It is extremely difficult to maintain a cold rotating structure in a geometrically thick state. Various processes have been evoked to account for these observations, such as supernova heating \citep{Wada02}, outflowing winds \citep{Elitzur06} and warped discs \citep{Lawrence07}. 

The orientation of the narrow line region, as delimited by the ionization cones and primarily collimated by the torus, is essentially uncorrelated with the galaxy disc \citep{Fischer13}. The same lack of correlation between the jet and the torus must tell us something about the dynamics of the innermost region, close to the SMBH, and about the fueling process itself. The reasons given to explain why this misalignment exists include: minor mergers of gas-rich dwarf galaxies; alignment of the jet with the BH spin, leading to the conclusion that the spins of the central BHs are uncorrelated with the rotation of the circumnuclear discs; and the fact that on scales of 10 pc the turbulent movement of clumps can fuel the AGN in discrete events with randomly oriented orbits. The last option provides a different explanation for fractions of type 1 and type 2 AGNs \citep{Lawrence10}, with type 2 AGNs having a larger misalignment. 

Within the central hundreds of parsecs of a galaxy, most of the gas is in the molecular phase and its morphology and dynamics have been studied mainly through the 2-1 and 1-0 lines of $^{12}$CO with resolution $<1^{\prime\prime}$, which corresponds to $<10-50$ pc (see the NUGA project by \citealt{Garcia03}), and through H$_2$ ro-vibrational transitions, which trace the warm gas (T$\approx 700-2000$ K), with similar resolution. Close to the nucleus, 90\% of the AGNs show H$_2$ molecular emission \citep{Ardila04} distributed in disc-like structures; and down to a radius $<25$~pc, the ratio between rotational velocity and velocity dispersion $V_{rot}/\sigma~=~0.8\pm0.3$ \citep{Muller13}, which means that in this region the gas kinematics is dominated by turbulent motions in a thick disc. It is important to emphasize that the torus is different from the $\sim30$ pc scale distribution of the observable thick discs, which has a small covering factor of $\sim1\%$. It must be closer to the nucleus to account for the required obscuration \citep{Hicks09}.

\begin{figure*}
 \begin{minipage}{0.65\columnwidth}
  \resizebox{\hsize}{!}{\includegraphics{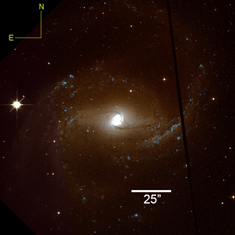}}
 \end{minipage}
 \begin{minipage}{0.65\columnwidth}
  \resizebox{\hsize}{!}{\includegraphics{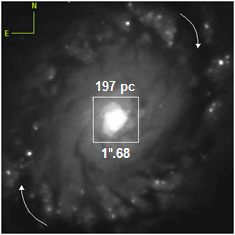}}
 \end{minipage}
 \caption{Left: HST image of NGC 6951 in the \textsc{F658N} (blue)+\textsc{F814W} (white) filters. Right: HST image in the I filter with 9 arcsec$^{2}$. The white square denotes the FOV (field of view) of $1^{\prime\prime}.68\times1^{\prime\prime}.68$, dimensions that will be used in all images in this paper, centred in the bulge, and with North on top. The arrows show where the large scale bar connects with the stellar ring.}
 \label{fig:6951structure}
\end{figure*}

NGC 6951 is a barred spiral galaxy SAB(rs)bc \citep{Vaucouleurs91} (Fig.~\ref{fig:6951structure}) and, according to the classification proposed by \citet{Veilleux87}, hosts a Seyfert 2 AGN \citep{Ho95, Ho97112}, although \citet{Perez00} argued that its nucleus can be considered as a transition object between a high excitation LINER and a nitrogen overabundant Seyfert 2. It is at a distance of 24.1 Mpc \citep{Tully88} ($1^{\prime\prime}=117$ pc) and has a disc inclination of $i=46.2$\textdegree~\citep{Haan09}. This galaxy has a starburst ring connected to an outer bar, with spiral dust lanes connected to the nucleus and a typical outflow in the form of two ionization cones \citep{Thaisa07}.

\citet{Haan09} noted the absence of neighboring galaxies within a projected distance of 1 Mpc and redshift differences lower than 500 km s$^{-1}$, suggesting there has been no external gravitational influence for the last $\sim10^9$ years. These authors also showed that the nucleus of NGC 6951 is HI deficient, implying that its ISM is dominated by the molecular phase. \citet{Krips07} reported that the nuclear gas of NGC 6951 has properties similar to the Seyfert galaxies NGC 1068 and M51. Both objects have high central HCN-to-CO ratios, suggesting that NGC 6951 might display the same scenario whereby the molecular gas chemistry is dominated by X-ray radiation (NGC 1068: \citealt{Usero04}; M51: \citealt{Matsushita98}).

A radio compact nuclear component was reported by means of VLA observations by \citet{Saikia02}, with an angular size of $\sim0^{\prime\prime}.7\times\sim0^{\prime\prime}.2$, corresponding to $\sim80\times20$ pc$^{2}$ in the galaxy, and a position angle of 156\textdegree. The existence of jets in LLAGNs was studied by \citet{Mezcua14}, who analyzed sub-arcsec archival data from VLA  and VLBA (very long baseline array) observations of eight nearby LLAGNs. They found that all these galaxies have pc-scale or larger radio jets. The non-detection of large-scale jets in this sample of galaxies suggests that this component is confined to a small region, either because of its considerable misalignment with the internal disc or because of its interaction with the ISM. 

In this work, we will analyze data from the Gemini North Telescope obtained with the Near-Infrared Integral Field Spectrograph (NIFS), in the K band, which allows us to study the warm molecular gas through the H$_2$ lines. This analysis is complemented with archive data obtained from the GMOS spectrograph and the HST, allowing the study of ionized gas.

The structure of this work is as follows. Section~\ref{sec:2} presents the reduction and treatment of the NIFS and GMOS data cubes. Section~\ref{sec:data} discusses the properties of the emission lines of the molecular gas, its spatial distribution and its kinematics, as well as estimates of its temperature. Section~\ref{sec:hst} presents the same results for the ionized gas. We continue in Section~\ref{sec:stellar} to present the stellar content and its kinematics. In Section~\ref{sec:discussion} we discuss the results and, finally, in Section~\ref{sec:conclusions} we draw our conclusions.

\section{Observations, reductions and data treatment}
\label{sec:2}
\subsection{Near infrared data - NIFS}

The data presented here were obtained during the night of September 2, 2012, using the NIFS instrument \citep{McGregor03} on the Gemini North Telescope, operating with the adaptive optics module ALTAIR (Altitude conjugate adaptive optics for infrared), in Laser Guide Star (LGS) mode, under programme GN-2012B-Q-44. The pixel size of the instrument is $0^{\prime\prime}.103\times 0^{\prime\prime}.043$ in $x$ and $y$ directions, respectively, with a FOV of $\sim3^{\prime\prime}\times 3^{\prime\prime}$. The observations were made in the K band ($1.99-2.40~\mu$m), with a spectral resolution of $R\approx 5290$ ($\approx 30$ km $s^{-1}$), and consisted of 8 individual exposures alternating on-source and sky observations of 750 s each. However, only four of them, which presented best seeing $\sim0^{\prime\prime}.35$, were used.  

The data were reduced using tasks of the NIFS package in IRAF environment. The procedure included trimming the images, flat-fielding, sky subtraction, correcting for spatial distortions and wavelength calibration. We removed the telluric bands and calibrated the flux using the A0V star HIP 107555. This standard star was chosen because it has a sharper point spread function (PSF) than the other available one, in spite of the lack of flux on the blue part of the spectra, from $2.08-2.20~\mu$m. Since we are not interested in stellar population synthesis because of the short wavelength interval of the fit, we simply removed this defect by fitting a spline to the continuum, together with the stellar absorption bands, and subtracting the fit from the original spectra, keeping only the emission lines. 
At the end of the data reduction process, the IFU data cubes were generated by the \texttt{nifcube} task, which re-sampled them to spaxels of $\sim0^{\prime\prime}.05\times 0^{\prime\prime}.05$. It is important to mention that, after this procedure, the fluxes were no longer the same as in the previous data format (29 slices), and were corrected by multiplying the data cubes by a factor of 0.54.  

After the reduction, we performed a data treatment procedure described in more detail in \citet{Menezes14}. We corrected the Differential Atmospheric Refraction (DAR) empirically, fitting third degree polynomials through the spatial location of the centroids along the data cube, one for each spatial dimension, to maintain them at the same position in each wavelength. Although the DAR is small in the infrared, the high spatial resolution of NIFS observations, with adaptive optics (AO), can account for up to 3 spaxels of displacement due to this effect. At the end of the correction, all the centroids, measured from the peak in the image of the stellar continuum, remained the same with a precision of $0^{\prime\prime}.01$. This practical approach is the most precise to remove this effect, since the theoretical curves do not reproduce spatial displacements properly along the spectral axis. This is crucial to combine the data cubes.

The next step was the spatial re-sampling of the data, followed by a quadratic interpolation (Isquadratic), which was performed by fitting a quadratic function to each group of four adjacent spaxels along the lines and columns of the image. This procedure, which preserves the surface flux of the images, aims to improve the visualization of the contours of the structures. But when followed by the deconvolution process, the interpolation leads to better resolution. The new sampling was $0^{\prime\prime}.021\times 0^{\prime\prime}.021$, in keeping with the Nyquist criterion for re-sampling data of the smallest pixel size, corresponding to half of the sampling frequency of the image in the $y$ direction. It is worth noticing that this procedure introduces high spatial frequency components, which can be seen in the Fourier transform of the images. These components can be removed by the Butterworth spatial filtering in the frequency domain.

We combined the four data cubes, with four dithering points in both spatial dimensions, through a median. However, to apply a median, it was necessary to multiply the data cubes by a numerical factor because of the flux difference between the observations. The process eliminates the cosmic rays, as well as defects in the CCD. 
The total FOV was reduced to comprise only the region inside the stellar ring, resulting in a square of $\sim$200~pc$^{2}$ ($1^{\prime\prime}.68\times 1^{\prime\prime}.68$) with a final spectrum range from 2.1 to $2.4\mu$m.

\subsubsection{Spatial and spectral Butterworth filtering}

This process consists of filtering in the frequency domain and calculating the Fourier transform of the images or spectra in the data cube. This is followed by multiplying the Fourier transform by the image corresponding to the Butterwoth spatial filter \citep{Gonzalez02}, and calculating the inverse Fourier transform of this product. The idea is to remove the high frequencies by applying a low-pass filter and perform the inverse Fourier transform to the new filtered image/spectrum.

When spatial filtering the NIFS data, the most adequate procedure is the multiplication of a filter with elliptical shape by one with rectangular shape (see the mathematical definition in \citealt{Menezes14}), the main reason for this being the asymmetric shape of the pixels on the CCD. The best cut-off frequency found in NGC 6951 was \textsl{f}=0.35 for the \textsl{x} axis and \textsl{f}=0.4 for the \textsl{y} axis in both filters, where 1 corresponds to the Nyquist frequency for our data. We checked the final data cube to determine how much of the flux of the central region, with an aperture radius of $\sim0^{\prime\prime}.2$, changed after the filtering and found a variation of 2\%, which means that the PSF was practically not affected, considering the amount of noise removed from the data.

\begin{figure}
 \begin{minipage}{0.85\columnwidth}
  \resizebox{\hsize}{!}{\includegraphics{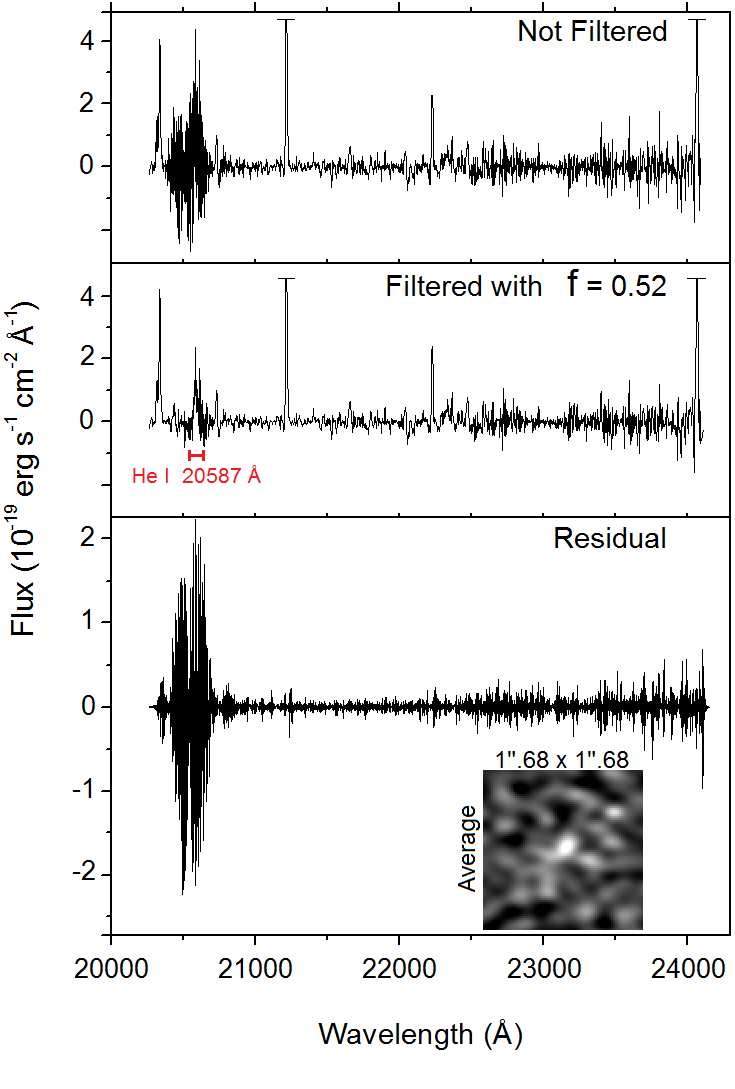}}
 \end{minipage}
 \caption{Top: average spectrum with an aperture radius of $\sim0^{\prime\prime}.2$. Middle: filtered spectrum with cut-off frequency of f=0.52, with the detection of the He I line. Bottom: the noise filtered and the average image of the data cube of noise.}
 \label{fig:butspec}
\end{figure}

Similarly, the filtering can be applied in one dimension, to each spectrum, with just one cut-off frequency, with \textsl{f}=0.52 in this case. We verified the flux with the H$_2$ $\lambda$21218 \AA~emission line for the same aperture and found that the difference between the filtered and non-filtered fluxes of this line was less than 3\%. Fig.~\ref{fig:butspec} shows that, in the filtered spectrum, it becomes possible to see the He I $\lambda$20585 \AA~line that was previously undetected due to the high level of noise. In the bottom of Fig.~\ref{fig:butspec}, we plotted the filtered noise and the corresponding average image. The data cube with spectral filtering was not used to extract the kinematics of the emission lines or the stellar kinematics, but only for Principal Component Analysis (PCA) tomography \citep{Steiner09} and for the construction of the He I images.

\subsubsection{Richardson-Lucy deconvolution}

The PSF can generally be well described by a Gaussian or a Moffat function. However, for AO corrected images, this approach is not valid because of the complex shape of the PSF. For observations with AO, one expects a PSF with two components - a central diffraction spike, given by the Airy function, and a surrounding halo, given by a Lorentzian function -, but with the available spatial pixelation it is not possible to resolve the Airy function and this component is well described by a Gaussian. However, one cannot assume such a simple scenario in the case of NIFS data cubes, because the AO correction introduces complex profiles in the PSF that are not well fitted by any combination of the functions mentioned above. Since there is no observable point-like source in the data cubes of NGC 6951, the alternative solution was to estimate the PSF from the standard star data cube used in the data reduction, taking a small interval over the continuum centred at 2.2 $\mu$m. The standard star has a seeing of $0^{\prime\prime}.44$. 

\begin{figure}
 \begin{minipage}{0.85\columnwidth}
  \resizebox{\hsize}{!}{\includegraphics{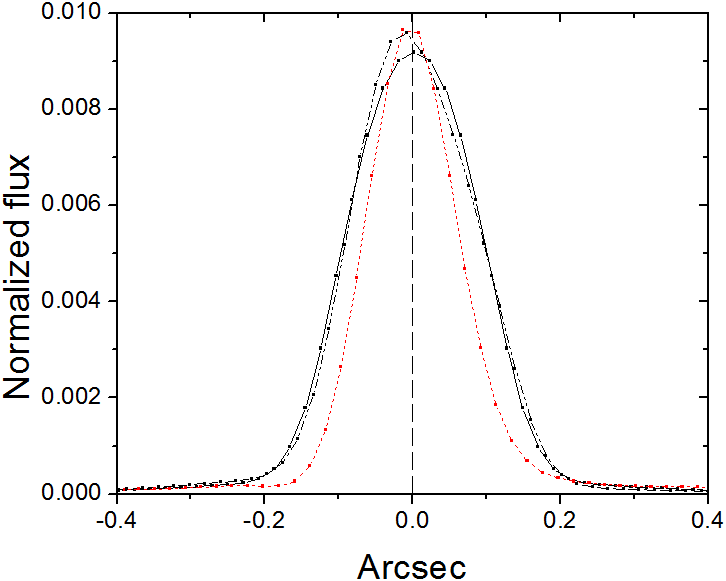}}
 \end{minipage}
 \caption{PSF profiles of the standard star along the \textsl{x} axis of the original image (black dashed-dotted curve) and after the symmetrization (black curve). The profile in the \textsl{y} axis remains unchanged (red curve). The FWHMs are $0^{\prime\prime}.17$, $0^{\prime\prime}.2$ and $0^{\prime\prime}.12$, respectively.}
 \label{fig:psf}
\end{figure}

As the NIFS spaxels have dimensions of $0^{\prime\prime}.103\times 0^{\prime\prime}.043$, the PSF of an individual star may become asymmetric in the \textsl{x}-axis (largest spaxels dimensions). The reason is that in this dimension the spaxel size is larger than the FWHM of the Airy profile in the K band ($\sim0^{\prime\prime}.06$), while in the \textsl{y}-axis it is smaller. Depending on the relative positioning of the star centroid on the spaxel, the lateral spaxels in the \textsl{x}-axis will be more or less illuminated. In order to minimize this effect in the galaxy exposures, we carefully chose the dithering positions to sample distinct parts on the spaxel. In the median image any asymmetry will mostly be removed. In the single observation of the standard star, however, one can see the asymmety in the peak of the PSF in the \textsl{x}-axis (Fig.~\ref{fig:psf}). As expected this is not noticed in the \textsl{y}-axis. Therefore we decided to symmetrize the PSF before applying the Richardson-Lucy deconvolution \citep{Richardson72, Lucy74}. This step was performed adding the same image of the PSF, inverted in the \textsl{x}-axis, to the original image before normalizing it. We have successfully applied Richardson-Lucy deconvolution with symmetric PSFs in a variety of similar situations (\citealt{MenezesTese,Menezes14,Menezes15,Menezes15b}).

\begin{figure*}
 \begin{minipage}{0.68\columnwidth}
  \resizebox{\hsize}{!}{\includegraphics{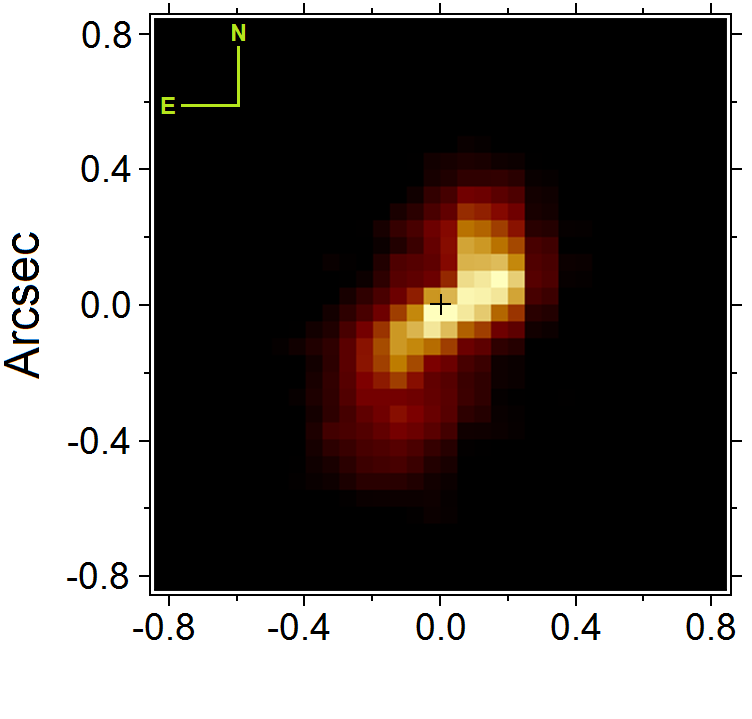}}
 \end{minipage}
 \begin{minipage}{0.55\columnwidth}
  \resizebox{\hsize}{!}{\includegraphics{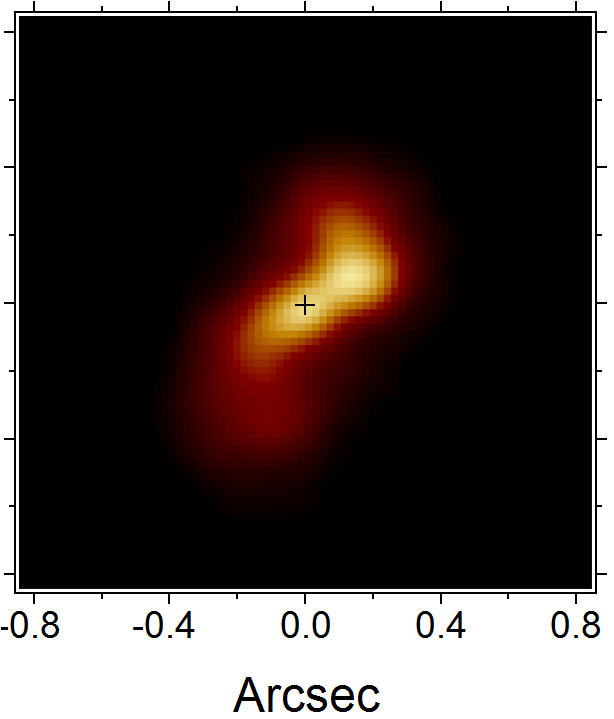}}
 \end{minipage}
\begin{minipage}{0.685\columnwidth}
  \resizebox{\hsize}{!}{\includegraphics{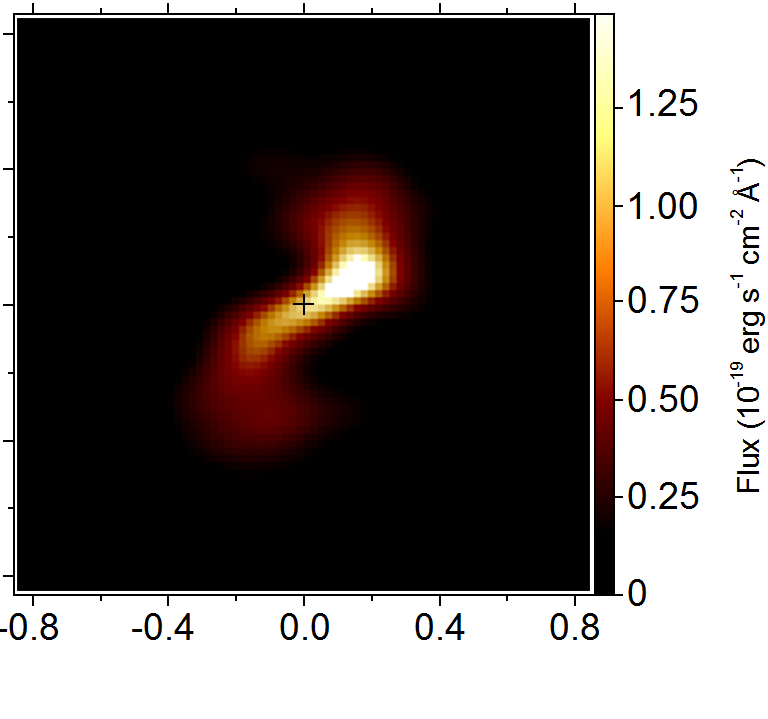}}
 \end{minipage}
 \caption{Average image of the continuum-subtracted data cube of the H$_{2}$ lines, shown after the reduction process with pixel scale of $0^{\prime\prime}.05$ (left); after the re-sampling to $0^{\prime\prime}.021\times 0^{\prime\prime}.021$ and the Butterworth filtering (middle); and after the Richardson-Lucy deconvolution process (right). The black cross denotes the centre of the bulge.}
 \label{fig:h2}
\end{figure*}

We applied six iterations to each image of the data cube of NGC 6951, which proved to be the best choice in this case. Assuming that the FWHM measured after the reduction and treatment of the standard star is the same for a point-like source in the galaxy, the spatial resolution, after the deconvolution applied to the galaxy data cube (based in the percentage of FWHM decreased in the continuum of the galaxy) was estimated in $\sim0^{\prime\prime}.09$. This corresponds to $\sim$5 pc in the galaxy, or 2 spaxels after the re-sampling. In Fig.~\ref{fig:h2}~the spatial structure for the H$_2$ $\lambda$21218 \AA~line is shown in three different stages of the data treatment, where can be seen new structures that had not been detected as clearly before.
 
\subsection{Optical data - GMOS}
\label{sec:datagmos}

The observations obtained with the Integral Field Unit of the Gemini Multi-Object Spectrograph (GMOS IFU) on the Gemini North telescope, on the nights of August 31, 2006 and September 1, 2006, were already analyzed and published by \citet{Thaisa07}. Here we re-analyze the same data, but take into account only three of the nine data cubes that are centred on the AGN of NGC 6951. The exposures of 500 s have a FOV of $5^{\prime\prime}\times 7^{\prime\prime}$ with a seeing of $\sim0^{\prime\prime}.5$ during the night, corresponding to a spatial resolution of $\sim40$ pc in the galaxy. The wavelength range is 5600-7000 \AA, with a spectral resolution of $R\approx 2300$ ($\sim130$ km $s^{-1}$).

The data reduction was made in IRAF environment, using the \texttt{gemini.gmos} package. The steps comprised bias subtraction, flat-fielding, correction of spatial distortions and wavelength calibration. At the end of the process, 3 data cubes were obtained with spatial pixels of $0^{\prime\prime}.1\times 0^{\prime\prime}.1$.
We then applied a data treatment very similar to the one applied to the NIFS data cubes, including the following steps: DAR correction, median of the cubes, Butterworth spatial filtering and, finally, Richardson-Lucy deconvolution, with the FHWM estimated from the calibration star. The process consisted of six iterations and a Gaussian PSF with FWHM=$0^{\prime\prime}.52$. The final PSF was estimated by convolving the HST image, in the V filter, with an Gaussian PSF. The FWHM of such PSF was chosen to result in a HST convolved image that has the same FWHM measured in the deconvolved GMOS data. Thus, the estimated PSF after the deconvolution, has the FWHM of the Gaussian used in this process, that is $0^{\prime\prime}.45$.

\section{Analysis and results for the emission lines: the molecular gas}
\label{sec:data}
\subsection{The H$_{2}$ lines}

\begin{figure}
 \begin{minipage}{0.95\columnwidth}
  \resizebox{\hsize}{!}{\includegraphics{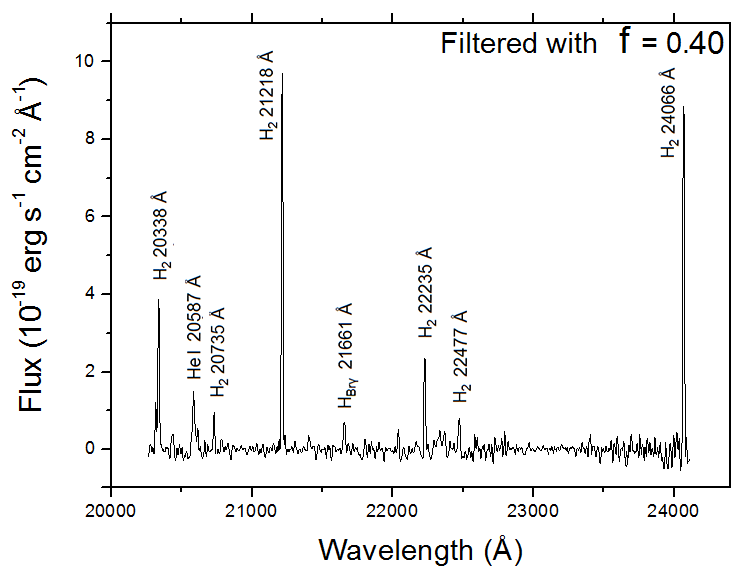}}
 \end{minipage}
 \caption{Average spectrum of the data cube of NGC 6951 after the continuum subtraction, within an aperture radius of $0^{\prime\prime}.2$, centred on the galaxy bulge.}
 \label{fig:molecular}
\end{figure}

In Fig.~\ref{fig:molecular}, we show the average spectrum of NGC 6951 in the K band, after the subtraction of the continuum, within an aperture radius of $0^{\prime\prime}.2$, centred on the bulge. In this spectrum, we can identify six H$_2$~ro-vibrational transitions, the Br$\gamma$~$\lambda$21661 \AA~and the HeI $\lambda$20585 \AA~lines. The continuum was subtracted by a simple spline function fitting, masking the emission line regions, and later subtracting it from the original spectra. We did not use a stellar population synthesis to perform this subtraction because of the inaccurate fits given for such a short wavelength interval and of the poor spectral resolution compared with the data.

\begin{figure*}
\resizebox{0.80\hsize}{!}{\includegraphics{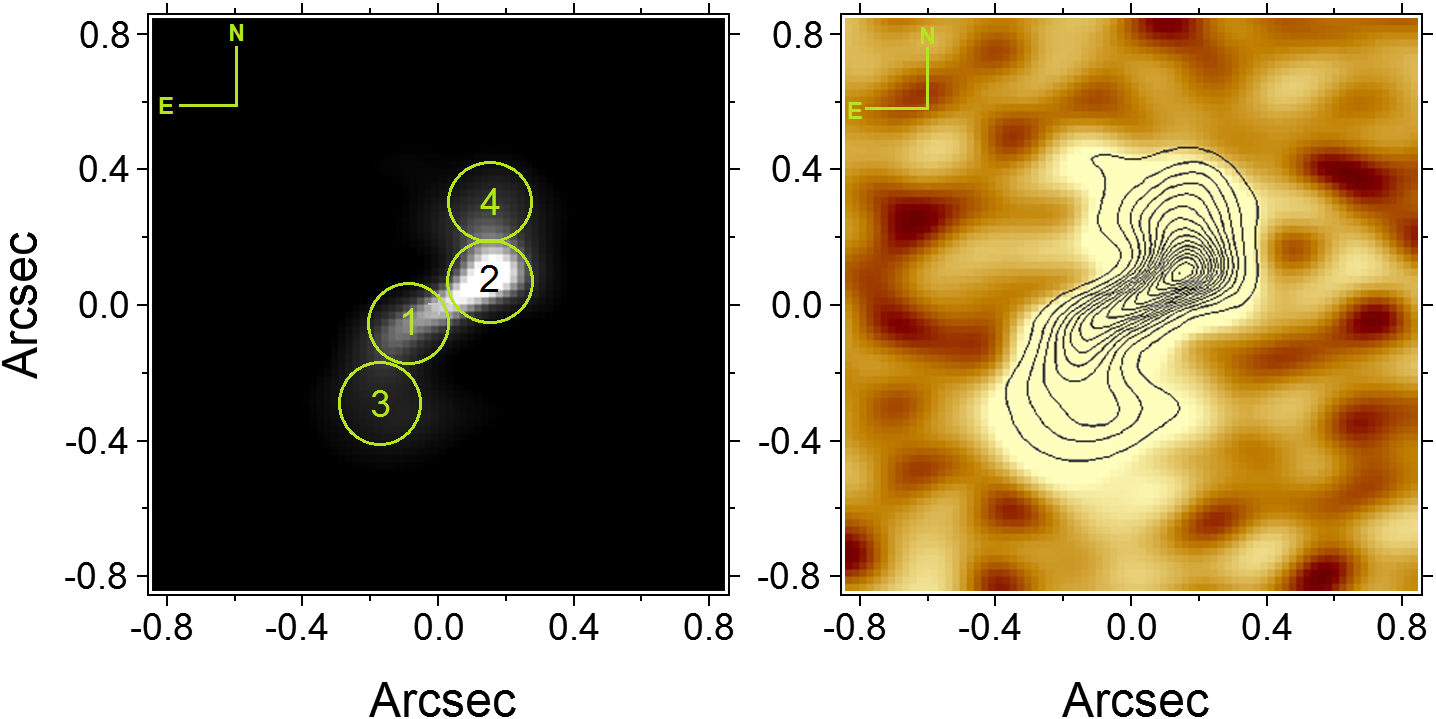}}
\caption{Left: the molecular gas image, showing the four regions with radii of $0^{\prime\prime}.1$, from which the line fluxes were extracted. Right: The same image, scaled to show the noise fluctuation, and the area of significant H$_{2}$ emission. The contours are in squared scale, with the inner contour corresponding to 15\% and the outer contour to 1\% of the flux peak.}
 \label{fig:h2flux}
\end{figure*}

In Fig.~\ref{fig:h2}, the images were extracted from the continuum-subtracted data cube with the He I and Br$\gamma$ lines masked, so that all the images from the NIFS data cube will represent only the molecular gas emission (referred to simply as ``molecular gas''), except where we specify individual line images. In the right panel of Fig.~\ref{fig:h2}, one can see that the image of the molecular gas presents a flat structure, possibly an edge-on disc, with PA=124\textdegree, with its extremities extending almost perpendicularly to the flat structure, in opposite directions. Table~\ref{table:flux}~displays the measured fluxes for the H$_{2}$~emission lines of the spectra extracted from four circular regions, with radii of $0^{\prime\prime}.1$, at the positions identified in Fig.~\ref{fig:h2flux}~(left panel), and also the total flux of the spectrum of a circular region, with a radius of $0^{\prime\prime}.5$, centred on the bulge. Due to the large calibration uncertainties, the errors in the absolute fluxes are $\sim$30\%~or even higher, as pointed out by the multiplication factors applied prior to calculating the median between the four data cubes. 

As indicated in Fig.~\ref{fig:h2flux}~(left), regions 1 and 2 are along the PA=124\textdegree, while 3 and 4 represent the faint emission at the elongated ends. According to \citet{Thaisa07}, the near side of the galaxy is to the southwest and the far side to the northeast, but it is interesting to note, when comparing the relative intensities of the lines between regions 3 and 4, that they are only slightly different (an average of 11\% more intense in region 4), suggesting that there is no appreciable variation in dust extinction. In Fig.~\ref{fig:h2flux}~(right panel), we present the same image in order to show the noise level and, thus, the region from where we can safely extract the properties of the molecular gas and where we can make reliable fits.

\begin{table*}
\begin{center} 
\caption[linefluxes]
{Measured H$_{2}$ emission line fluxes for the four circular regions marked in Fig.~\ref{fig:h2flux}, with radii of $0^{\prime\prime}.1$ and total emission within an aperture radius of $0^{\prime\prime}.5$, including Br$\gamma$ and He I lines. All values are in units of $10^{-16}$~erg s$^{-1}$~cm$^{-2}$. Errors do not exceed the typical 30\% of uncertainty in flux calibration.}
\begin{tabular}{ccccccc}
\hline \hline
$\lambda_{vac}$~(\AA) & ID & 1 & 2 & 3 & 4 & Total \\ \hline
20 338 & H$_{2}$ 1-0 S(2) & $2.94\pm0.19$ & $2.39\pm0.22$ & $1.43\pm0.12$ & $1.80\pm0.09$ & $12.72\pm0.44$ \\
20 585 & HeI & -- & -- & -- & -- & $5.48\pm0.51$ \\
20 735 & H$_{2}$ 2-1 S(3) & $0.78\pm0.09$ & $0.67\pm0.04$ & $0.34\pm0.03$ & $0.46\pm0.09$ & $3.91\pm0.58$ \\ 
21 218 & H$_{2}$ 1-0 S(1) & $6.15\pm0.15$ & $6.40\pm0.14$ & $3.93\pm0.24$ & $4.24\pm0.12$ & $51.81\pm1.89$ \\
21 661 & Br$\gamma$ & -- & -- & -- & -- & $4.04\pm0.76$ \\
22 235 & H$_{2}$ 1-0 S(0) & $1.66\pm0.17$ & $1.84\pm0.12$ & $1.04\pm0.14$ & $1.33\pm0.13$ & $11.20\pm1.26$ \\
22 477 & H$_{2}$ 2-1 S(1) & $0.65\pm0.13$ & $0.50\pm0.14$ & $0.50\pm0.10$ & $0.42\pm0.06$ & $4.78\pm1.57$ \\
24 066 & H$_{2}$ 1-0 Q(1) & $5.78\pm0.76$ & $6.86\pm0.42$ & $3.58\pm0.13$ & $3.94\pm0.35$ & $43.39\pm5.13$ \\
\hline 
\end{tabular}
\label{table:flux} 
\end{center} 
\end{table*}

\subsection{H$_{2}$ kinematics}

\begin{figure*}
 \resizebox{0.99\hsize}{!}{\includegraphics{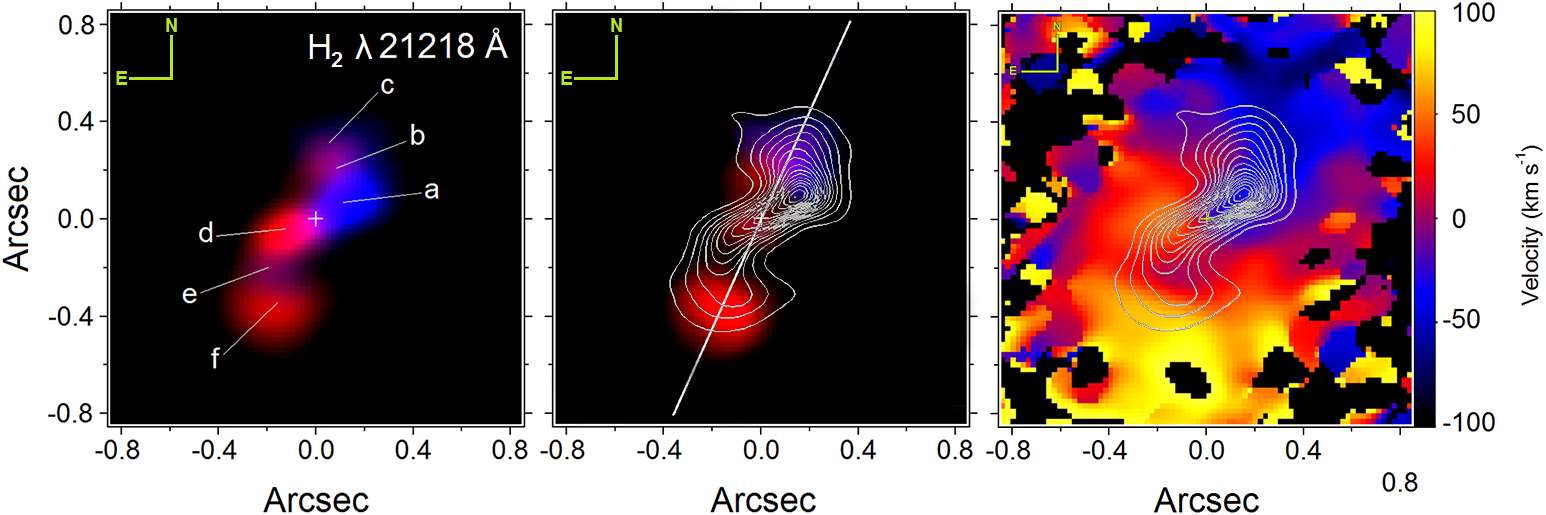}}
 \caption{Left: image of the H$_{2}$ $\lambda$~21218 \AA~line, in blueshift (from -226 km $s^{-1}$ to 0 km $s^{-1}$) and redshift (from -0 km $s^{-1}$ to 226 km $s^{-1}$). The letters indicate the positions of the circular regions, with radii of $0^{\prime\prime}.1$, from which the line profiles were extracted, and are shown in Fig.~\ref{fig:6regionsprofiles2}. Middle: the same line for velocities $v<-140$~km s$^{-1}$~and $v>140$~km s$^{-1}$~(that is, the blue and the red wings only). The white line shows the PA$_{radio}$=156\textdegree~and the contours represent the molecular gas. Right: radial velocity map with one Gaussian fit for the H$_{2}$ $\lambda$~21218 \AA~line at each spaxel.}
 \label{fig:6regionsprofiles1}
\end{figure*}

\begin{figure*}
\resizebox{0.85\hsize}{!}{\includegraphics{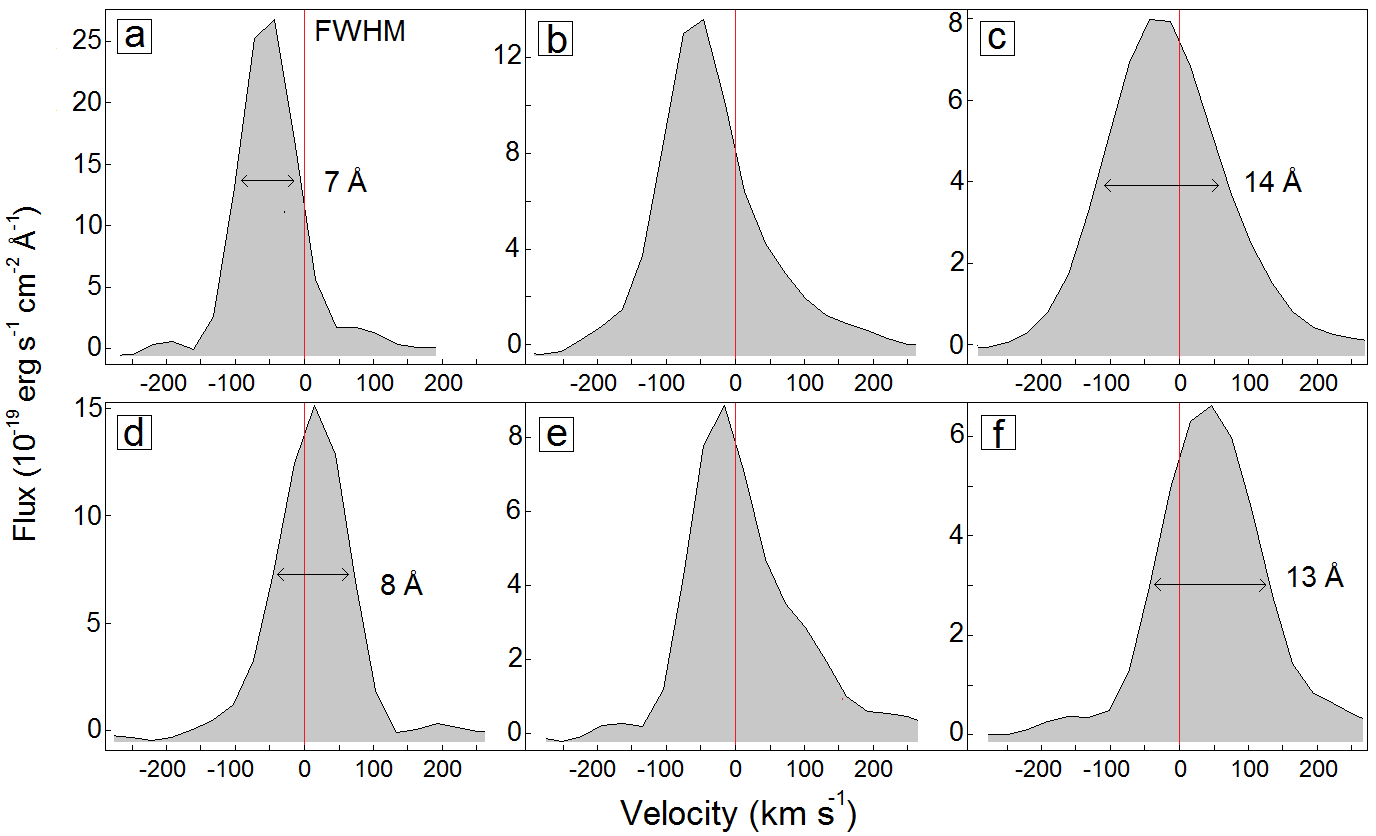}}
\caption{The H$_{2}$ $\lambda$21218 \AA~line profiles for the six marked regions in Fig.~\ref{fig:6regionsprofiles1}. The vertical red line denotes the rest frame wavelength, with zero velocity.}
 \label{fig:6regionsprofiles2}
\end{figure*}

Fig.~\ref{fig:6regionsprofiles1} (left panel) allows us to visualize the complete spatial distribution of the red and blue wings for the H$_{2}$ $\lambda$21218 \AA~line and the image of the wings of the same line with velocities $v<-140$~km s$^{-1}$~and $v>140$~km s$^{-1}$ (middle panel), as well as the contours of the molecular gas. Despite differences due mainly to the signal-to-noise ratio (hereafter S/N), all the H$_{2}$ lines have consistent kinematics. The six different regions denoted by letters correspond to the locations from where we extracted the spectra of circular regions with radii of $0^{\prime\prime}.1$, to determine the H$_{2}$ $\lambda$21218 \AA~line profiles. All the line profiles are shown in Fig.~\ref{fig:6regionsprofiles2}, with the upper section showing the blueshifted FWHM, increasing for regions more distant from the centre, with similar behaviour for the redshifted lines in the lower section. Regions $b$~and $e$~have clearly asymmetric profiles and may represent the transition between two velocity regimes, one more turbulent and dominated by velocity dispersion and the other seen in the narrow line profiles located in the disc. In Fig.~\ref{fig:6regionsprofiles1} (right panel), we show the radial velocity map based on a single Gaussian fit to the H$_{2}$ $\lambda$21218 \AA~line at each spaxel, with the contours of the molecular gas. The fits beyond a radius of $\sim0^{\prime\prime}.4$ could lead to a misleading interpretation of the extended kinematic, where the S/N is too low for a reliable analysis (as shown in Fig.~\ref{fig:h2flux}, right panel). The kinematic axis associated with the disc is 103\textdegree$\pm3$\textdegree.

Our interpretation is that the narrow line profiles of regions 1 and 2 in Fig.~\ref{fig:h2flux} (left) represent the emission of a disc, with PA=124\textdegree~and a thickness barely resolved, giving an upper limit of $\sim$20 pc. This disc is connected to the turbulent gas associated with regions 3 and 4. The behaviour of the two velocity regimes may have originated from some interaction between the disc and the jet, accounting for the high-velocity dispersion, both for cold and hot molecular gas, characteristic of nuclear outflows \citep{Martin06, Davies14}. This hypothesis is strengthened by the orientation of the radio position angle, which agrees with the position angle of the turbulent gas (Fig.~\ref{fig:6regionsprofiles1} middle panel).

\begin{figure}
 \begin{minipage}{0.45\textwidth}
  \resizebox{\hsize}{!}{\includegraphics{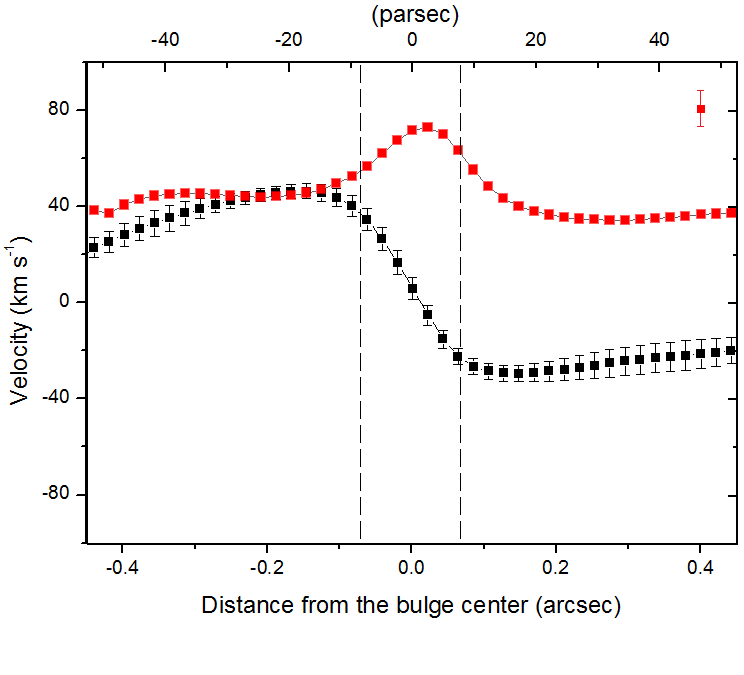}}
 \end{minipage}
 \begin{minipage}{0.45\textwidth}
  \resizebox{\hsize}{!}{\includegraphics{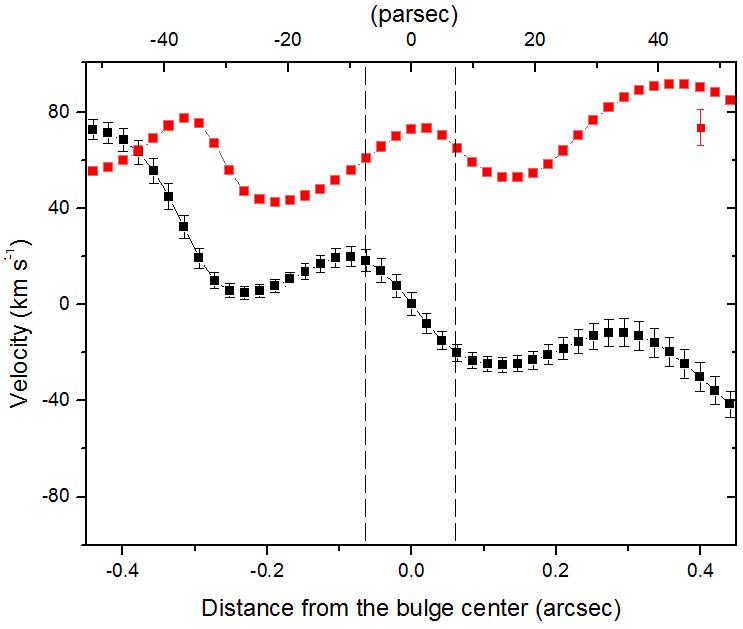}}
 \end{minipage}
 \caption{Radial velocity profile (black squares) and velocity dispersion (red squares) for the H$_{2}$ $\lambda$21218 \AA~line, measured along the position angles of 103\textdegree~(top), representing the molecular disc, and 156\textdegree~(bottom), the orientation of the radio emission. The vertical dashed lines denote the PSF of the standard star, and the error bars for the velocity dispersion, corrected for instrumental broadening, are shown within the graph.}
\label{fig:h2vel}
\end{figure}

The velocity profiles shown in Fig.~\ref{fig:h2vel} were extracted from two different orientations, corresponding to the disc and the radio emission position angles, respectively, with a pseudo-slit width equivalent to $0^{\prime\prime}.1$. The kinematic axis for the disc (103\textdegree$\pm3$\textdegree) is distinct from what we measured in the H$_{2}$ $\lambda$21218 \AA~narrow line profiles image of the red and blue wings (the disc in Fig.~\ref{fig:6regionsprofiles1}, left panel), which was 124\textdegree$\pm6$\textdegree. This difference means that the maximum velocity does not coincide with the maximum flux. Indeed, the radial velocity profile displays the characteristic curve of a disc, which does not decrease so steeply over large distances. The radial velocity in the disc spans from +40 to -40 km s$^{-1}$ and the disc has an average velocity dispersion, corrected for the instrumental broadening, of 36 $\pm$4 km s$^{-1}$, similar to the radial velocity. The velocity profile along the orientation of the radio emission reaches $\sim$70 km s$^{-1}$ in the turbulent regions, which have an average velocity dispersion of 69 $\pm$2 km s$^{-1}$ in region 1, and 70 $\pm$2 km s$^{-1}$~in region 2. Considering that the disc is seen almost edge-on, the radial velocities are probably close to the real values. Assuming an inclination of $\sim$90\textdegree~for the H$_{2}$ disc, it is inclined $\sim$44\textdegree~vis-\`a-vis the stellar disc of the galaxy. The kinematic axis is lower than that of H$\alpha$ (PA=125$\pm$10), found by \citet{Thaisa07}, where they assume that the gas is in the galactic disc.

\subsection{PCA tomography of the molecular gas data cube}
\label{sec:pcah2}

PCA is a statistical technique used to extract information from a large amount of data by calculating the correlations between their variables. It is defined as an orthogonal linear transformation that brings the data to a new uncorrelated coordinate system arranged in such a way that the first of these coordinates (eigenvector E1) explains the highest fraction of data variance, the second eigenvector explains the second highest fraction and so on. PCA tomography \citep{Steiner09} is a method that applies PCA to data cubes, where the variables are spectral pixels and the observables correspond to the spaxels of the data cube. Since eigenvectors are obtained as a function of wavelengths, their correlations have a shape similar to the spectra and are therefore called eigenspectra. On the other hand, the projections of the observables on the eigenvectors are also images, indicating where the correlations take place on the spatial coordinates. To interpret the results, it is necessary to analyze simultaneously the eigenspectra and the tomograms. For some applications see \citet{Ricci11}, \citet{Schnorr11}, \citet{Menezes13} and \citet{Ricci14I}.

\begin{figure*}
 \begin{minipage}{0.88\textwidth}
  \resizebox{\hsize}{!}{\includegraphics{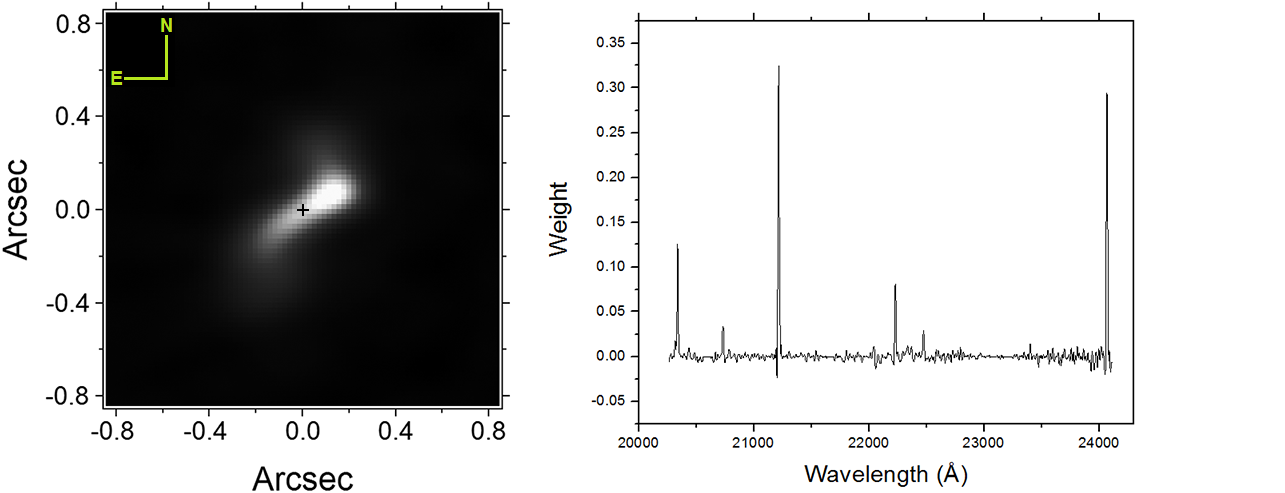}}
 \end{minipage}
 \begin{minipage}{0.88\textwidth}
  \resizebox{\hsize}{!}{\includegraphics{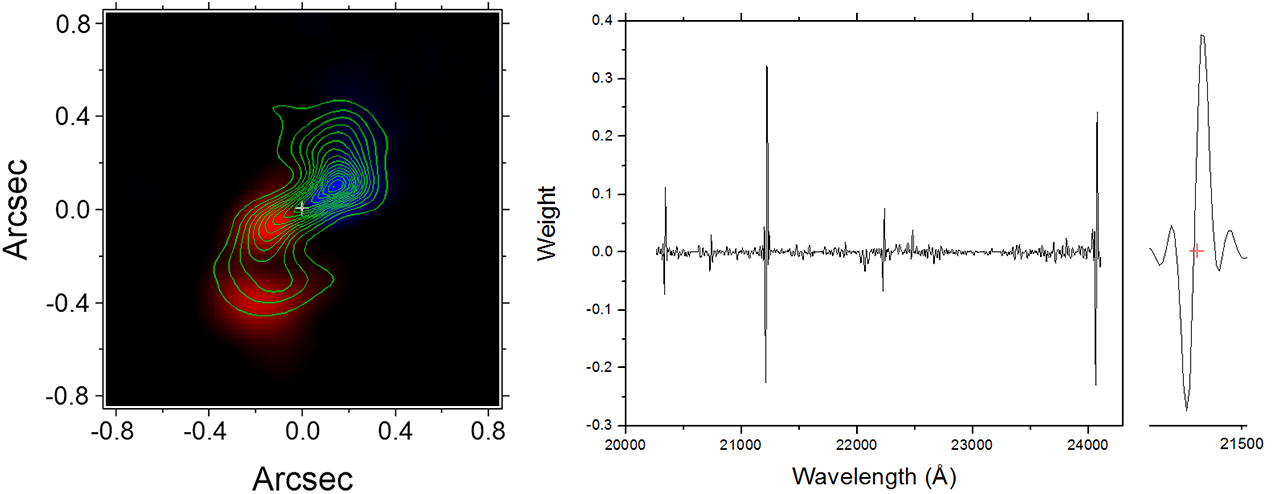}}
 \end{minipage}
\begin{minipage}{0.88\textwidth}
  \resizebox{\hsize}{!}{\includegraphics{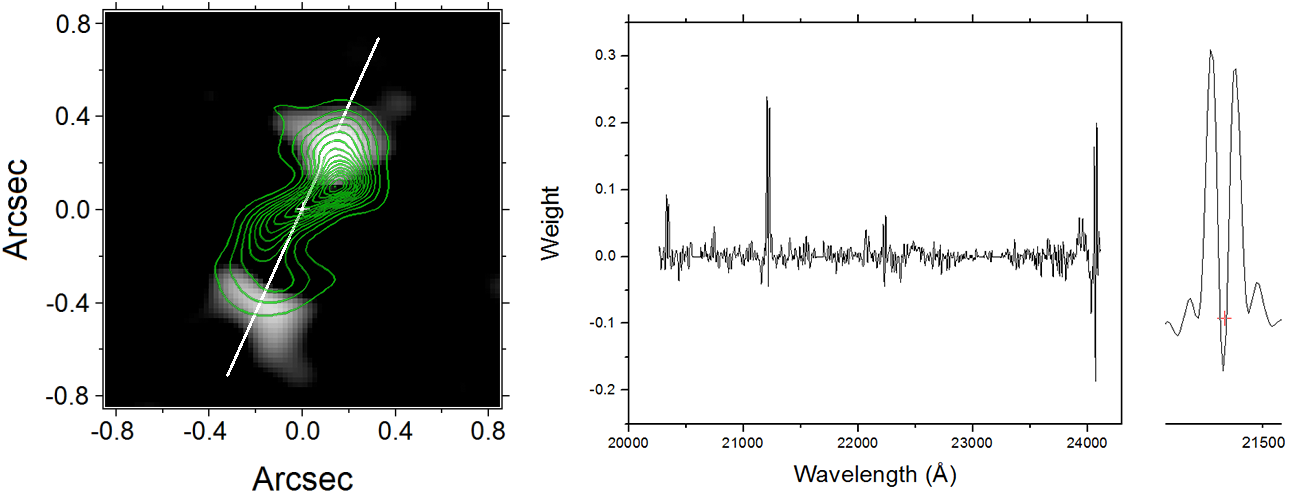}}
 \end{minipage}
 \caption{The first three tomograms and eigenspectra of NGC 6951, obtained with PCA tomography, from the NIFS data cube for the molecular gas. The middle panel shows the total negative weight (blueshifted wing) in blue, and the positive weight (redshifted wing) in red. For better visualization, the H$_{2}$ $\lambda$21218 \AA~wavelength is zoomed in. The contours represent the molecular gas, the white line is the jet PA of 156\textdegree~and the cross denotes the bulge centre.}
\label{fig:h2pca}
\end{figure*}

In order to derive only the correlations between the molecular lines, we masked the Br$\gamma$ $\lambda$21661 \AA~and HeI $\lambda$20585 \AA~recombination lines and applied PCA tomography to the masked data cube. The most significant eigenvectors obtained with this procedure are shown in Fig.~\ref{fig:h2pca}.

One can see that the first eigenspectrum and the respective tomogram are quite similar, respectively, to the average spectrum (Fig.~\ref{fig:molecular}) and to the image of the average molecular gas data cube (Fig.~\ref{fig:h2}~right panel). This is expected because eigenvector 1 explains most of the data variance (61.15\%), although the NW side of the structure appears to be more prominent here because the tomogram has a higher weight where there is more emission in the data cube, compared to the calculation of the average emission. This structure has the same PA=124\textdegree$\pm$6\textdegree of the average emission and its elongated extremities are almost perpendicular. 

In eigenspectrum 2 (9.78\% of the data variance), there are correlations between wavelengths corresponding to the red wings of all H$_{2}$ lines, which are anti-correlated to the wavelengths corresponding to the blue wings of these lines, indicating a kinematic phenomenon of the molecular gas. An analysis of the morphology of tomogram 2 reveals that the blueshifted part of the line is in the NW part of the structure and the redshifted one in the SE part, with the bulge center exactly in the middle. Looking at the tomogram, there is also the same anti-correlation associated with the turbulent regions. The colours in Fig.~\ref{fig:h2pca} (middle panel) should be taken as the blueshifted and redshifted velocities with respect to the LOS.

In eigenspectrum 3 (2.07\% of the variance), the broad wings of the lines are correlated and have a weak anti-correlation to the narrow part of the central peaks. Our interpretation is that the bright areas of the tomogram correspond to regions where the FHWM of the emission lines are broader and, therefore, where the higher values for the velocity dispersion are found. The spatial location for this correlation agrees with the previous image of the red and blue wings of the molecular lines (middle panel of Fig.~\ref{fig:6regionsprofiles1}).

The contours in Fig.~\ref{fig:h2pca} show how the kinematic information of the tomograms is spatially related to the molecular structure, represented by the average data cube of the molecular gas. The kinematics of tomogram 2 is associated both with the disc (rotation) and with the turbulent regions (outflow). Since PCA analysis produces new uncorrelated eigenvectors, the tomograms interpreted as gas kinematics are quite precise in representing the structures in different velocity regimes, because the resolution of the spectra does not allow us to properly select the wavelength intervals to discriminate between the displacement of the narrow line profile and the increase of the FWHM in the same direction. In this case the PCA is a good way to show the spatial location of the turbulent gas. 

\subsection{Physical conditions of the molecular gas}

The H$_{2}$ emission comes from rotational and vibrational transitions with $\Delta J = -2, 0, +2$, where odd rotational states have parallel spins (\textsc{ortho}-H$_{2}$) and even J states have anti-parallel spins (\textsc{para}-H$_{2}$). For non-thermal excitation, followed by radiative decay, the ratio between lines from \textsc{ortho} molecules is constant (0.5-0.6 for the 2-1S(1)/1-0S(1) lines), but the \textsc{ortho-para} ratio is not. However, for thermal excitation, the \textsc{ortho-para} ratio is expected to be constant and the 1-0S(0)/1-0S(1) ratio is $\sim3$. 
 
These lines can be excited in two ways: by a non-thermal process, through fluorescence by UV photons \citep{Black87} and by a thermal process, produced either by X-ray \citep{Maloney96} or by shock heating \citep{Hollenbach89}. The temperature ranges from 514 K, for the first pure translational transition J(2-1), to $\sim4000$ K, when the molecules begin to be quickly destroyed by energetic collisions. The typical thermal value, where the excitation temperature is the same as the kinetic temperature, is $\sim 2000$ K, with critical densities of about $10^{5-6}$ cm$^{-3}$.

In Table~\ref{table:flux}, we show the fluxes, non-corrected for extinction, of the detected emission lines. They were measured for four different regions (Fig.~\ref{fig:h2flux} left). Some useful line ratios are shown in Table~\ref{table:fluxratio}. 
To remove the effect of the \textsc{ortho/para} ratio, \citet{Mouri94} compares different excitation models using the intensity ratios of 2-1 S(1)/1-0 S(1) (which occur for \textsc{ortho} molecules) and 1-0 S(2)/1-0 S(0) (for \textsc{para} molecules), where non-thermal values are expected to be constant. 
The line ratios for NGC 6951 suggests that its nucleus has thermal excitation and that regions 2, 3 and 4 lie closest to the theoretical point of the shock heating process \citep{Brand89}, establishing a possible distinction from region 1, which is compatible with X-ray excitation \citep{Lepp83}. The gas interaction with the shock driving source may be more evident in region 2, which is brighter than region 1.

\subsubsection{H$_{2}$ Population Diagram}
\label{sec:PD}

For the high density gas, where the collisional excitation and de-excitation are dominant, the relative populations of the ro-vibrational levels $n_{\nu J}$ are described by the Boltzmann distribution. Given the relative level of the H$_{2}$ transitions, one may calculate the ratio of different population densities, which are proportional to observed column densities, versus the energy of the upper level, in what is called a population diagram. For a thermally excited gas, all the transition values lie on a straight line and the corresponding slope is inversely proportional to the gas temperature.
The column densities can be derived with the formula

\begin{equation}
N(\nu,J)=\frac{f}{A(\nu J, \nu^\prime J^\prime)} \times~\frac{\lambda}{hc} \times \frac{4\pi}{\Omega_{aper}}
\label{N}
\end{equation}

where $f$ is the measured flux, $A(\nu J, \nu^\prime J^\prime)$ is the transition probability from the $(\nu J)$ to the $(\nu^\prime J^\prime)$ state (taken from \citet{Wolniewicz98}), $\lambda$ is the rest frame wavelength, $h$ is the Planck constant, $c$ is the speed of light and $\Omega_{aper}$ is the aperture. 
By taking the logarithm of the ratio for two Boltzmann populations as a function of the column density, normalized by the transition $(\nu J)$ = $(1,3)$ (corresponding to the 1-0 S(1) line), we obtain the following equation

\begin{equation}
ln\frac{N(\nu J)/g_{J}}{N(1,3)/g_{3}}=\frac{-E(\nu J)/k}{T}+Constant
\label{population}
\end{equation}

\begin{figure}
\resizebox{0.95\hsize}{!}{\includegraphics{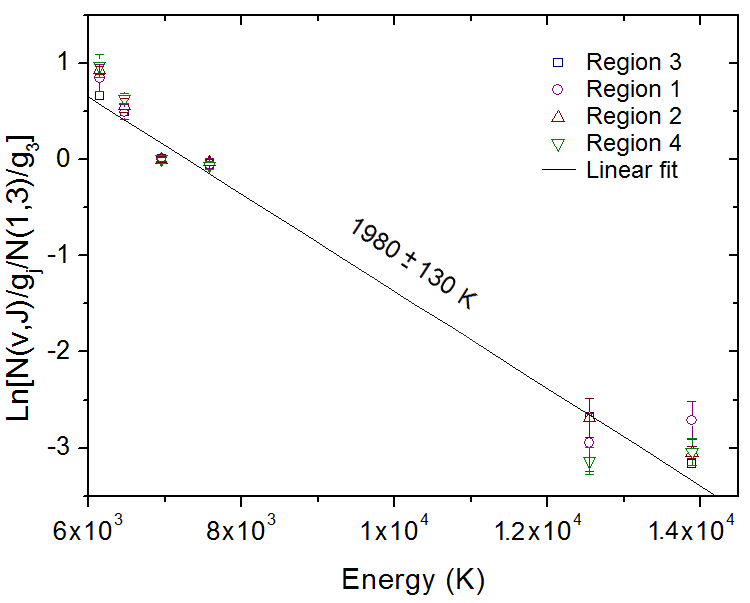}}
\caption{H$_{2}$ level population diagram relative to 1-0 S(1) for the four regions (marked in Fig.~\ref{fig:h2flux} left) of NGC 6951. The straight line is the linear fit with the slope consistent with an isothermal population at 1980 $\pm130$ K.}
 \label{fig:populationdiagramh2}
\end{figure}

The constant is independent of the transition. 
Fig.~\ref{fig:populationdiagramh2} shows the H$_{2}$ population diagram for the four regions (shown in Fig.~\ref{fig:h2flux} left), with the column densities normalized by the (1,3) transition, versus the energy of the upper level, in Kelvin degrees. A linear fit was performed  and we found a temperature of 1980 $\pm130$ K, compatible with thermal equilibrium for the H$_{2}$ gas.

According to excitation models described in \citet{Mouri94}, the 0.30 for the 1-0 S(0)/1-0 S(1) line ratio is $\sim$36\% above what would be expected for shock and X-ray excitation ($\sim$0.22), and closer to UV models. Due to the absence of a starburst in the nucleus and a resulting temperature incompatible with excitation by stars, we discarded UV excitation. Supernova remnants have a typical value of $\sim$0.22 for the 1-0 S(0)/1-0 S(1) line ratio \citep[excitation by shocks,][]{Mouri94}, lower than the ratio of $\sim$0.30 measured for NGC 6951. We argue that, taking into account the geometry of the molecular structure, it is very unlikely that this ratio is due to supernova remnants but, instead, might still be due to shock events, emanating from another source of energy. Similar ratios were found in the literature for several galaxies, attributed to shock excitation or present in very disturbed systems, e.g., 0.28$\pm$0.05 for NGC 520, a merging system \citep{Kotilainen01}; 0.29 for NGC 660, with two inclined dust lanes \citep{Schinnerer02}; 0.24$\pm$0.01 for NGC 1266, probably exited by C-shocks \citep{Pellegrini13}; 0.27$\pm$0.01 for NGC 1275, compatible with shock-excitation and turbulent heating \citep{Scharwachter13}; 0.3$\pm$0.1 for NGC 5929 \citep{Bower93}; and 0.18 for the galaxy Arp102B \citep{Stauffer83}, where the last two ratios have a strong cloud-jet interaction.

The alternative for a supernova would be the presence of a jet, where a relativistic plasma continuously hits the molecular gas. In fact, the nucleus of NGC 6951 has a central radio emission, which is slightly eccentric, suggesting that the radio jet could be stopped by the dense distribution of gas.

\begin{table*}
\begin{center} 
\caption[linefluxes]
{Measured H$_{2}$ line ratios for the four circular regions marked in Fig.~\ref{fig:h2flux}, with radii of $0^{\prime\prime}.1$, and the ratio for the total emission, within an aperture radius of $0^{\prime\prime}.5$. The ratio 1-0 S(1)/Br$\gamma$~is also shown.}
\begin{tabular}{cccccc}
\hline \hline
Line ratio & 1 & 2 & 3 & 4 & Total \\ \hline
$\frac{2-1 S(1)}{1-0 S(1)}$ & $0.10\pm0.02$ & $0.08\pm0.02$ & $0.13\pm0.03$ & $0.10\pm0.02$ & $0.09\pm0.03$ \\ \\
$\frac{1-0 S(0)}{1-0 S(1)}$ & $0.27\pm0.03$ & $0.29\pm0.03$ & $0.27\pm0.05$ & $0.31\pm0.04$ & $0.30\pm0.04$ \\ \\
$\frac{1-0 S(2)}{1-0 S(0)}$ & $1.77\pm0.29$ & $1.30\pm0.19$ & $1.37\pm0.18$ & $1.37\pm0.27$ & $1.14\pm0.20$ \\ \\
$\frac{1-0 S(1)}{Br\gamma}$ & $9.4\pm1.25$ & $9.5\pm1.9$ & -- & -- & $12.8\pm2.9$ \\
\hline 
\end{tabular}
\label{table:fluxratio} 
\end{center} 
\end{table*}

\section{Analysis and results for the emission lines: the ionized gas}
\label{sec:hst}
\subsection{HST images: the ionization cone and interstellar extinction}

\begin{figure*}
\resizebox{0.99\hsize}{!}{\includegraphics{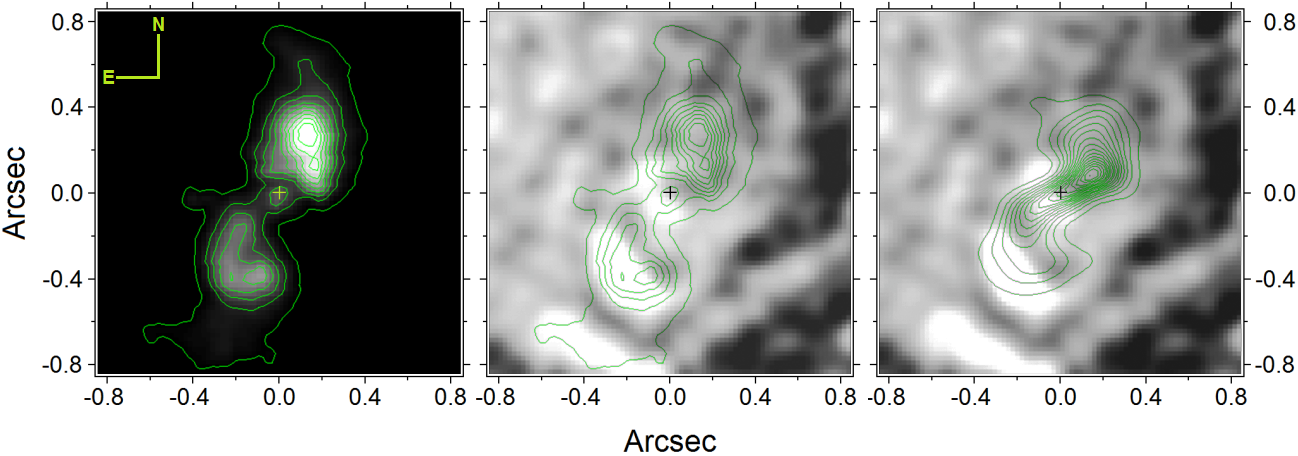}}
\caption{HST images. Left: (H$\alpha$+[N II])/I image. The contours are in linear scale, with the inner contour corresponding to 17\% and the outer contour to 3\% of the flux peak. Middle: (V-I) image with the contours of the previous image. Right: (V-I) with the contours of the molecular gas image. Darker regions correspond to higher extinction.}
 \label{fig:hst}
\end{figure*}

We analyzed images obtained with the Hubble Space Telescope (HST), retrieved from the HST archive, with the filters \textsc{F814W} (I band) and \textsc{F658N} (H$\alpha$+[N II]) from the instrument ACS WFC1, with scale of $0^{\prime\prime}.05$ per pixel, and with the \textsc{F547M} (V band) filter, obtained with the WFPC2/PC with scale of $0^{\prime\prime}.046$ per pixel. In the panels of Fig.~\ref{fig:hst}, we highlight the structure of the ionized gas and show the image of (V-I) with the contours of the ionized and molecular gas. The darker regions in the (V-I) images correspond to higher extinctions, which is in agreement with the structure map based likewise on HST images shown by \citet{Thaisa07} (see their Fig.4), where the SW region of the nucleus is clearly more obscured.

The first image in Fig.~\ref{fig:hst} (representing the H$\alpha$+[N II] emission) shows a double structure, symmetric with respect to a weak central point-like emission. We interpret this double structure as two ionization cones, the NW component being twice as intense as the SE one, which displays an arc shape. The line connecting the ionization cones has a PA=153\textdegree$\pm$2\textdegree. The central point-like weak emission coincides with the peak of the stellar bulge emission, seen in the I band. Hereafter, we assume that the bulge, seen in the I and K bands, has the same centre, which also defines the AGN position. It is worth noticing that there is only a weak emission in the vicinity of the AGN. The (V-I) image (Fig.~\ref{fig:hst}, middle panel) indicates where the extinction is higher (darker regions), which we attribute to the presence of dust. If we compare this image with the structure of the ionized gas, we see no correlation. This comparison shows that the NW side of the cone is intrinsically brighter than the SE side, as also seen in the NIFS data (Fig.~\ref{fig:h2}, right panel).
In Fig.~\ref{fig:hst} (right panel) we also see no correlation between the denser dust distribution and the molecular gas. This may appear contradictory, since H$_{2}$ molecules are formed on the surface of dust grains; however, once formed, the gas could be heated up to temperatures above the sublimation temperature of the grains, of $\sim$~1500 K. We calculated this temperature in the previous section as being T=1980 K and, in fact, this may be the case. Therefore, the dust may only be associated with the distribution of the cold molecular gas. We also noticed there is no clear evidence of any structure indicating spiral arms on these scales.

\begin{figure*}
\resizebox{0.75\hsize}{!}{\includegraphics{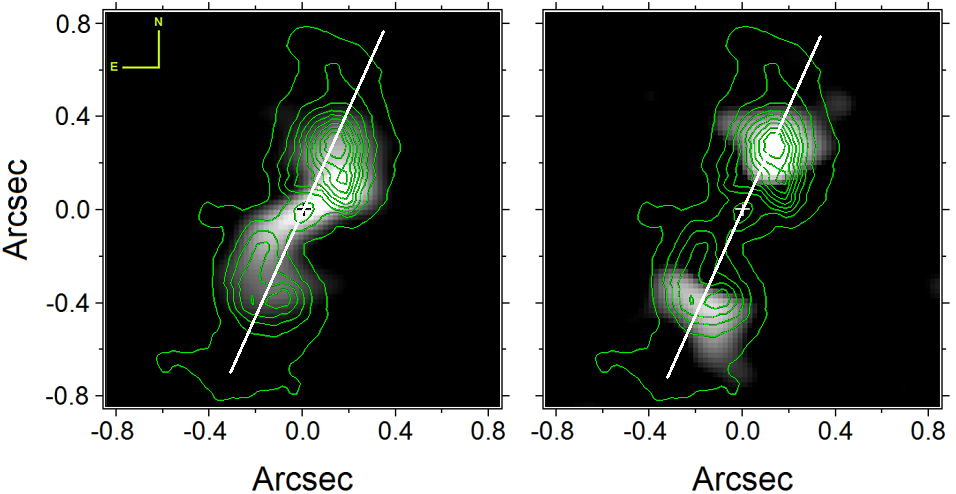}}
\caption{Left: image of the molecular gas (white) and the (H$\alpha$+[N II])/I image (green contours). Right: tomogram 3 (white) obtained with PCA tomography of the molecular gas data cube, with the same contours of the ionized gas, indicating where velocity dispersion is higher. The white line is the PA$_{Radio}$=156\textdegree.}
 \label{fig:hsth2}
\end{figure*}

By overlapping the contours of the ionized gas (as seen in the image from the HST) with the image of the molecular gas (Fig.~\ref{fig:hsth2} left), it becomes evident that the ionization cones are misaligned with the molecular gas in such a way that they coincide mainly with the H$_{2}$ emission that is not distributed along the disc, which is suggestive of some kind of interaction. The same can be said about the orientation of the radio emission, which coincides with that of the cones. The right image of Fig.~\ref{fig:hsth2} shows again the ionization cones and tomogram 3 of the NIFS data cube, which we interpreted as representing the regions where the gas is more turbulent, and they are co-spatial with the extremities of the ionized gas and have the same orientation of the radio jet. This scenario suggests that the high-velocity molecular gas, seen together with the ionized gas, could represent a nuclear outflow originated from the interaction between the jet and the molecular disc, in line with the analysis of the GMOS data in Sect~\ref{sec:gmos}.

\subsection{Ionized gas with NIFS: the Br$\gamma$ and He I emissions}

\begin{figure}
\resizebox{0.90\hsize}{!}{\includegraphics{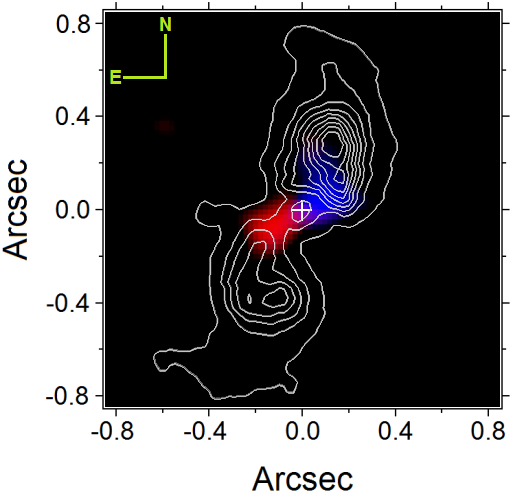}}
\caption{Image of the blue and red wings of the Br$\gamma$ emission line with the (H$\alpha$+[N II])/I image (white contours).}
 \label{fig:brg}
\end{figure}

We detected a weak emission of Br$\gamma$ (Fig.~\ref{fig:brg}), extending from the centre and connecting with the structure of the H$\alpha$+[N II] emission, with intensity and kinematic position angle consistent with those found for the H$_{2}$ lines (compare to Fig.~\ref{fig:6regionsprofiles1}, left panel).

This emission seems to fill the gap between the nucleus and the elongated ionization cones, suggesting that the emission seen in the HST image comes mostly from [N II]. This is confirmed by \citet{Thaisa07}, who found [N II]/H$\alpha=4-5$ for the nuclear region. In Sec.~\ref{sec:lr} we resume this discussion.

Examining the image from the blue and red wings of the line profile, we see that the kinematic centre of Br$\gamma$ also agrees well with the adopted centre of the AGN, as the centre of the bulge in the K band.
We measured the radial velocity, by fitting a Gaussian function to the Br$\gamma$ emission line in the spectra of regions 1 and 2 (left panel of Fig.~\ref{fig:h2flux}), and found a range from -40 km s$^{-1}$ to +40 km s$^{-1}$. This is the same as the extracted velocities for the molecular gas and also similar to the radial velocity measured for the ionized gas (Sect~\ref{sec:ionizedkin}). For the velocity dispersion, we measured an average of 55 $\pm9$ km s$^{-1}$ for region 1 and 80 $\pm36$ km s$^{-1}$ for region 2, which presents a very faint emission. These values are higher than those measured for the molecular gas in the disc (36 $\pm4$ km s$^{-1}$), although the errors are significantly larger.
Since this emission coincides spatially with the molecular disc, the hypothesis that it originates from the ionization cones is less favorable, given the extension of the ionized gas. 

The high value of $\approx 10$ for the H$_2$ $\lambda$21218 \AA/Br$\gamma$ ratio, shown in Table~\ref{table:fluxratio}, is well above the values found in literature, and far from the typical values of $\lesssim$0.6 for starburst galaxies (\citet{Mazzalay13}, \citet{Ardila04}, \citet{Ardila05}). In fact, this galaxy shows no indication of any significant star formation occurring in its nucleus \citep{VanderLaan13}. As we increase the aperture radius, the ratio increases as can be seen comparing the total ratio with regions 1 and 2. This difference is justified because the H$_{2}$ lines present a more extended emission. 

\begin{figure*}
\resizebox{0.90\hsize}{!}{\includegraphics{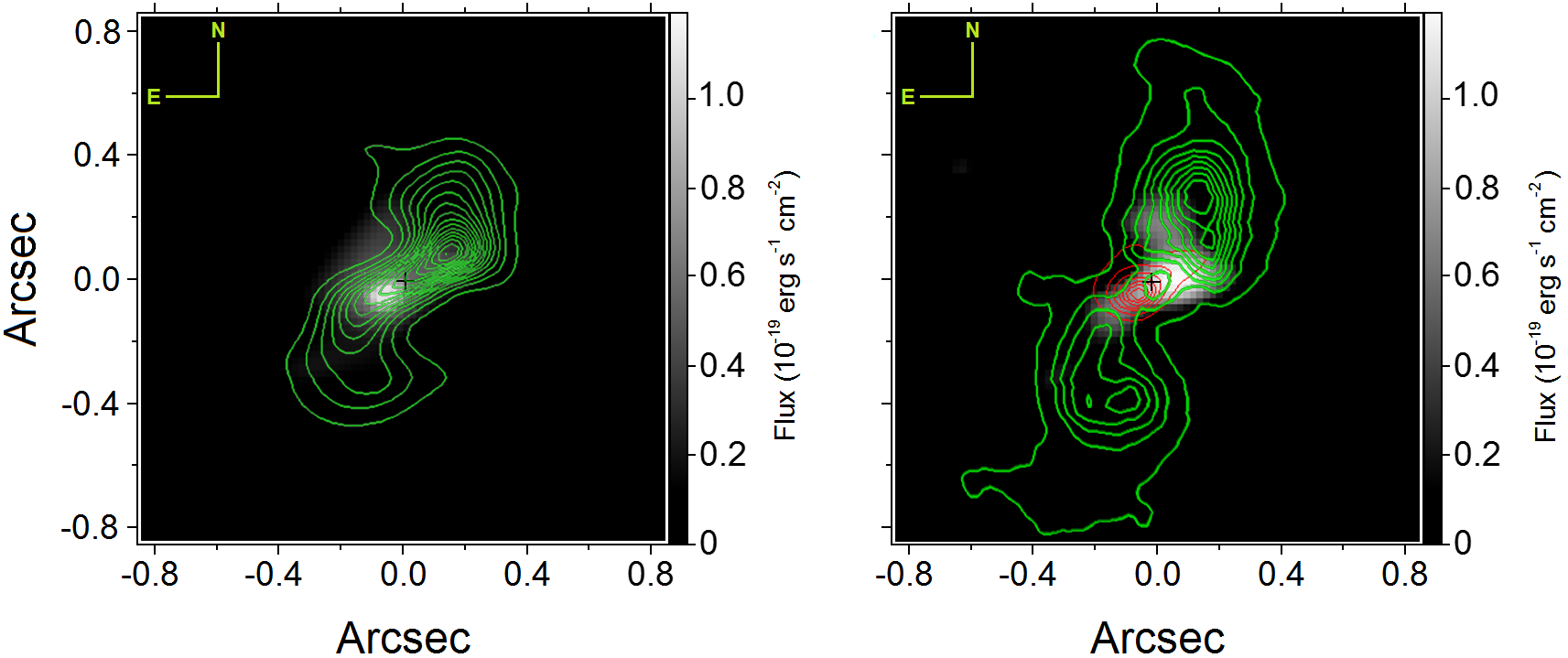}}
\caption{Left: image of the molecular gas (green contours) and the He I emission (white). Right: He I (red contours), Br$\gamma$ (white) and (H$\alpha$+[N II])/I (green contours). The crosses denote the centre of the bulge.}
 \label{fig:he}
\end{figure*}

In Fig.~\ref{fig:he}, we show the image of the integrated flux of the He I emission, which is asymmetric compared to the distribution of the molecular gas and with respect to the most intense Br$\gamma$ emission. Thus, we have a larger flux for the He I line in the SE part of the nucleus and the opposite case for the Br$\gamma$ emission. We emphasize that the He I line is located in a very noisy interval of the spectrum (Fig.~\ref{fig:butspec}), and even after filtering the high frequency noise, the remaining line profile may be affected by atmospheric absorption features.

\subsection{The GMOS data cube}
\label{sec:gmos}

In order to analyze the gas in the optical part of the spectrum, we took only the region inside the ring, corresponding to the same FOV and orientation used to analyze the NIFS data, and again subtracted the stellar emission. However, in this case, we performed a stellar population synthesis in each spectrum of the data cube, using the STARLIGHT Software \citep{Cid05} and the observed base of \citet{Bruzual03}. Stellar synthesis provides synthetic stellar spectra, from which a stellar data cube is generated. The stellar data cube is then used to subtract the stellar contribution from the total emission in the original data cube, as well as the contribution of the dust and the featureless continuum, to obtain the data cube of the emission lines. 

\begin{figure*}
\resizebox{0.85\hsize}{!}{\includegraphics{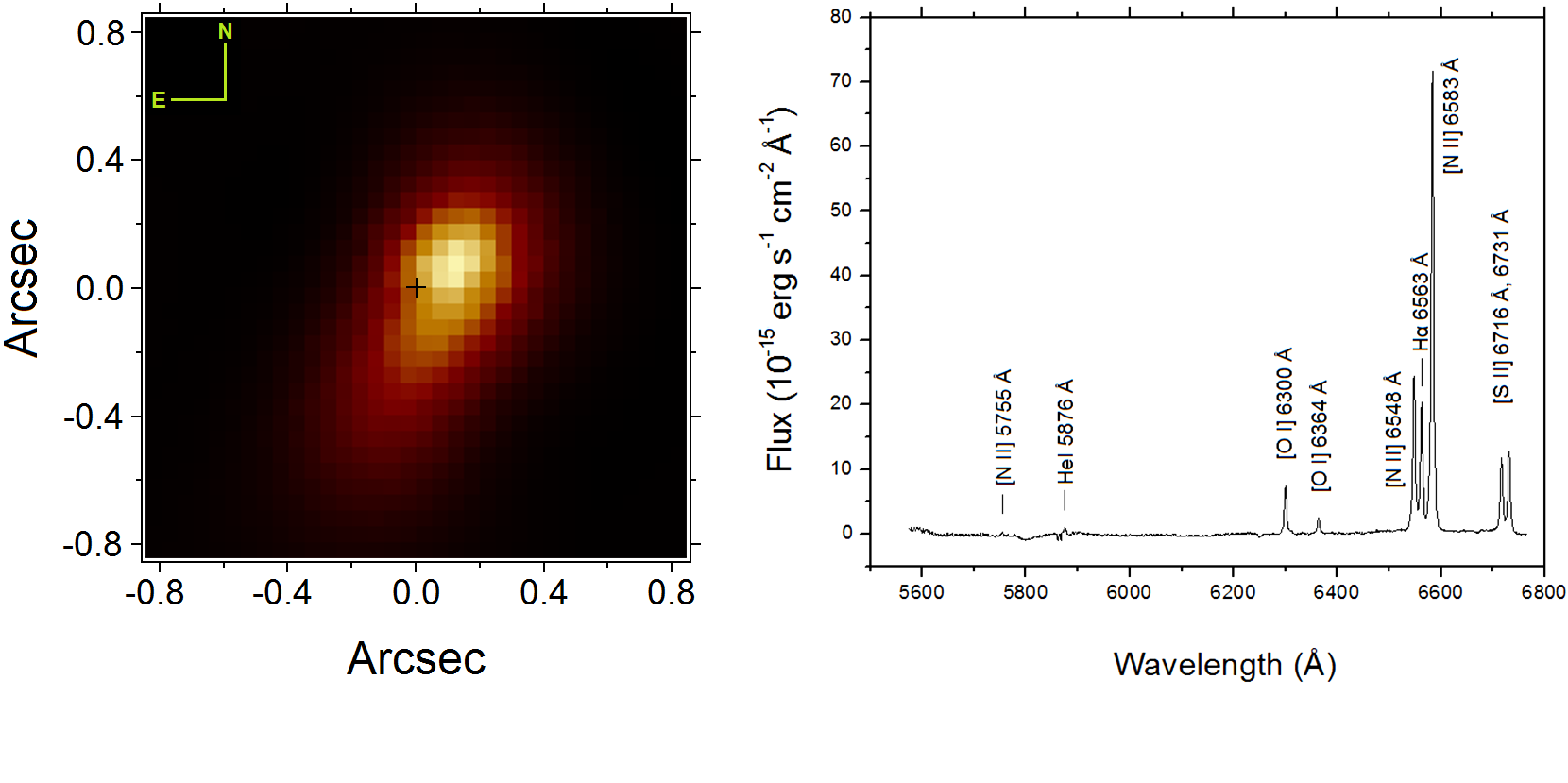}}
\caption{Left: the image shows the sum of the continuum-subtracted emission lines in the optical, from the GMOS data. The cross denotes the maximum continuum emission on the measured bulge centroid. Right: the corresponding continuum-subtracted spectrum extracted within an aperture radius of $0^{\prime\prime}.4$.}
 \label{fig:gmos}
\end{figure*}

Fig.~\ref{fig:gmos} shows the average image of the continuum-subtracted GMOS data cube, comprising only the emission lines, with the image centred on the bulge. We opted to show the average image for two reasons: to compare with PCA tomography results (next section) and because all the detected emission lines present quite similar spatial distribution and, therefore, similar images. We can only notice differences in spatial distributions among the detected emission lines taking their line ratio, as shown in Fig.~\ref{fig:NIIHa} (top and bottom panels). 

The centre of the bulge in the optical (Fig.~\ref{fig:gmos}) is marked with a cross and we see that the distribution of the ionized gas is clearly asymmetric with respect to it. One possible explanation for the displaced distribution of the integrated flux of the ionized gas with respect to the centre is the asymmetric distribution of dust in the nucleus. This is confirmed by Fig.~\ref{fig:hst} (middle panel), where the right side of the image is more susceptible to dust extinction, an effect that can be pronounced when the resolution is 6 times lower. In this case the centre of the bulge in the optical would be shifted to the left. 

To check if the asymmetric distribution of dust might displace the photometric centre of the GMOS continuum image, we measured the new centroid of the convolved HST image in the V filter (with a Gaussian PSF with FWHM=$0^{\prime\prime}.45$, as argued in Sec.~\ref{sec:datagmos}), and found a shift to the east of $0^{\prime\prime}.05$, i.e, one pixel in the GMOS data cube, and less than one pixel to the south. Although we can measure this displacement, this value is very sensitive to the FWHM of the estimated PSF. This is the reason why we do not consider this effect in the images derived from the GMOS data.      

A crucial aspect of this paper is choosing a reference frame to compare the images obtained with three instruments, namely, the NIFS and GMOS on Gemini North and the HST, and we emphasize that the respective centroids were taken as the bulge centre for each band, which we assume to coincide. This is a consistent choice, since there are good indications, in all observations, that the AGN centre is located in the centre of the bulge, as can be seen by the H$_{2}$ and Br$\gamma$ kinematic centre; the tomogram 2 of PCA tomography (for the GMOS data, in next section), and the punctual H$\alpha$+[N II] emission.  

\subsection{PCA Tomography of the GMOS data cube for the emission lines}

\begin{figure*}
 \begin{minipage}{0.75\textwidth}
  \resizebox{\hsize}{!}{\includegraphics{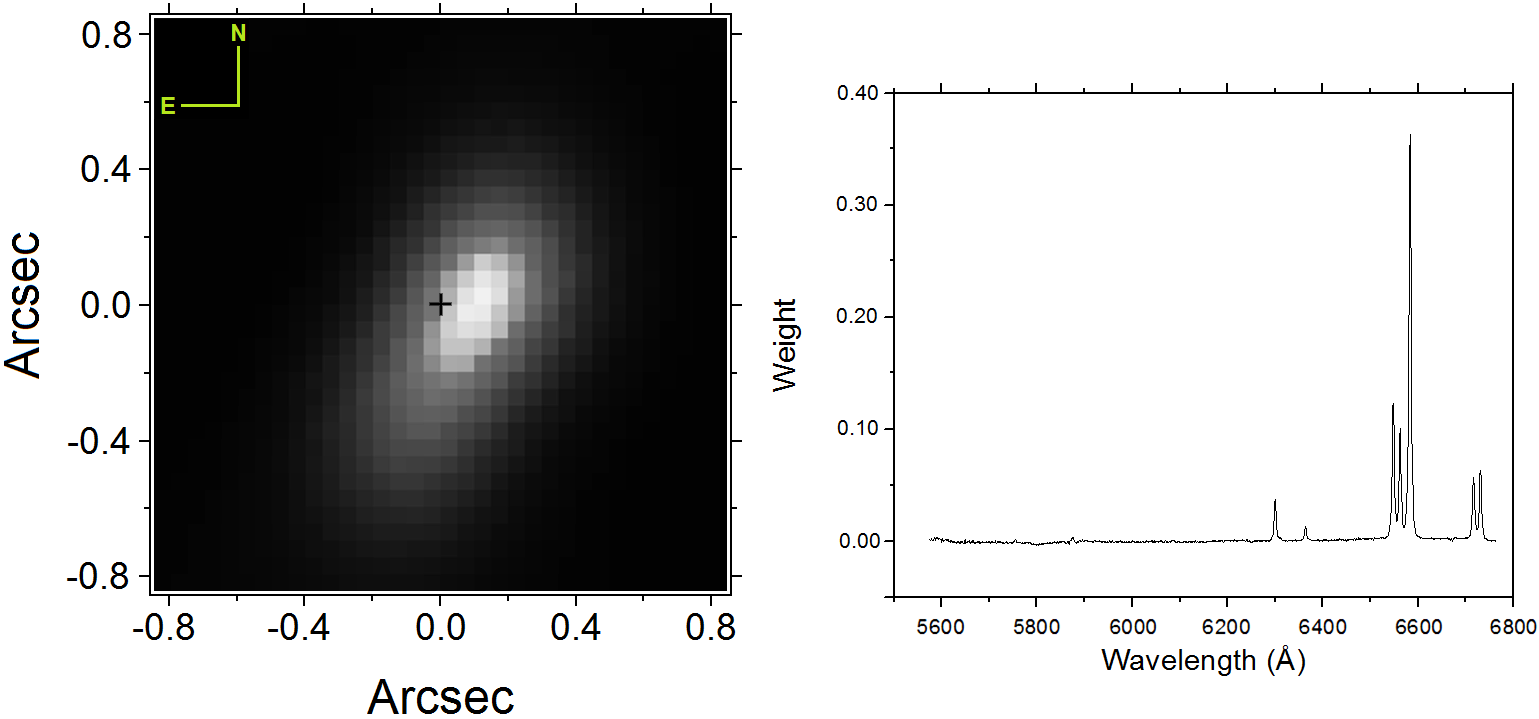}}
 \end{minipage}
 \begin{minipage}{0.75\textwidth}
  \resizebox{\hsize}{!}{\includegraphics{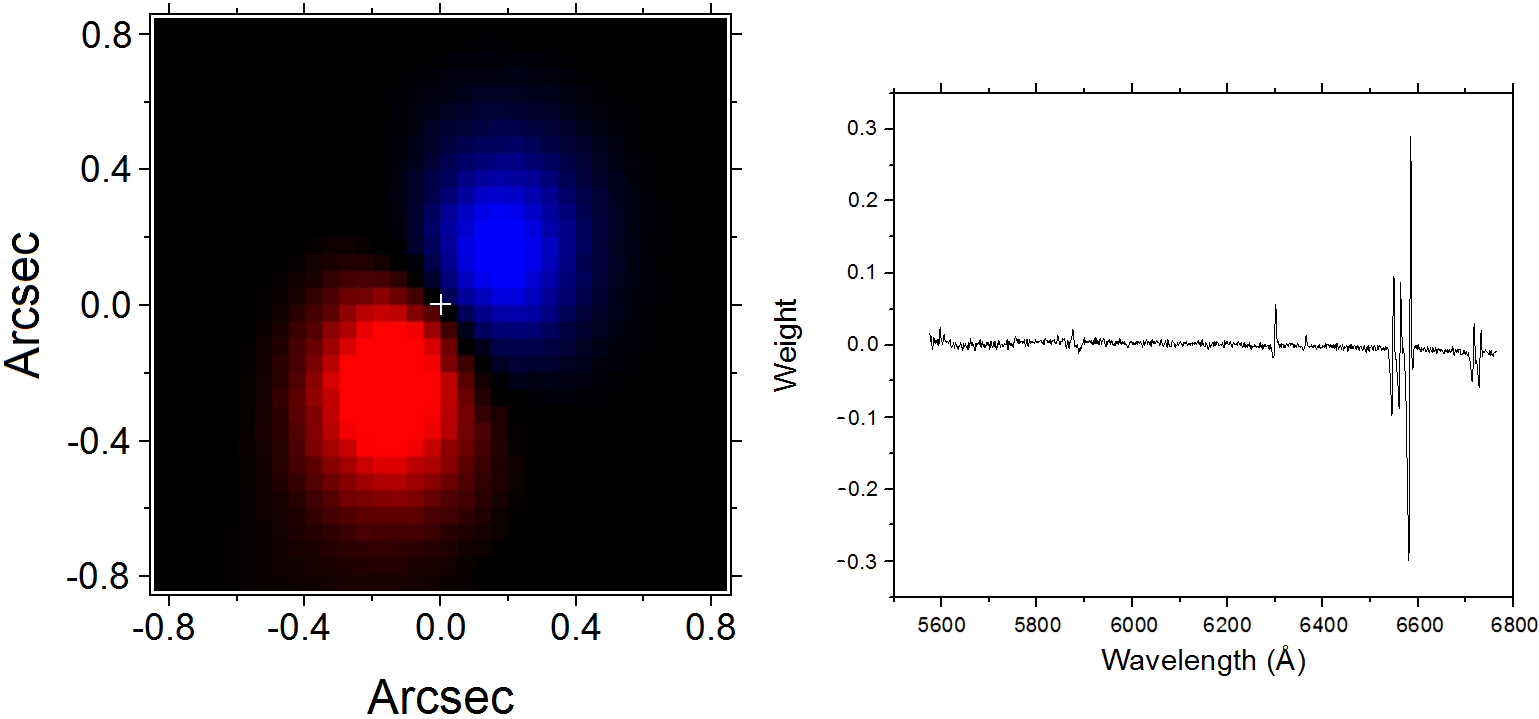}}
 \end{minipage}
\begin{minipage}{0.75\textwidth}
  \resizebox{\hsize}{!}{\includegraphics{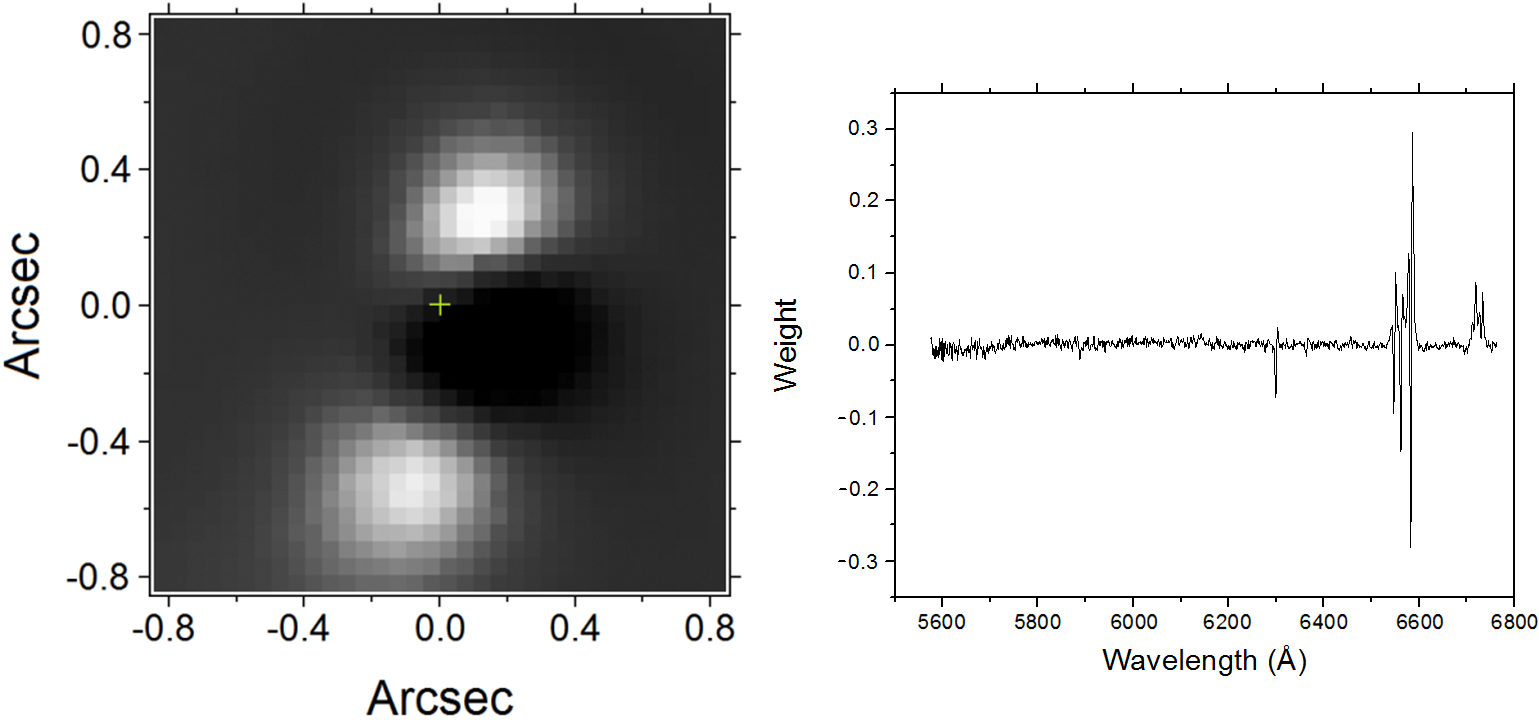}}
 \end{minipage}
 \caption{The first three tomograms and eigenspectra obtained with PCA tomography of NGC 6951, from the GMOS data cube for the ionized gas. The middle panel shows the total negative weight (blueshifted wing) in blue, and the positive weight (redshifted wing) in red.}
\label{fig:gmospca}
\end{figure*}

As in Sec.~\ref{sec:pcah2}, we applied PCA tomography to the GMOS gas data cube, and the first three eigenspectra and tomograms are presented in Fig.~\ref{fig:gmospca}.
The first eigenspectrum and tomogram provide basically the same information given by the average image and spectrum shown in Fig.~\ref{fig:gmos}, as they correspond to 88.39\% of the total variance. The second eigenspectrum (7.65\% of the variance) shows an anti-correlation between the blue and red wings of all emission lines, implying a kinematic phenomenon of the gas. Since we defined the structure depicted in the HST image as two ionization cones, with a similar orientation vis-\`a-vis the second tomogram, we interpreted this tomogram as a kinematic indication of the two ionization cones seen nearly edge-on, with PA=144\textdegree $\pm$3\textdegree. The NW cone is in the near side, above the galactic plane.
Eigenspectrum 3 shows similar correlations between the broad wings of the emission lines and anti-correlations between the narrow profile of the same lines, which are also seen in eigenspectrum 3 obtained with PCA tomography of the molecular gas data cube. Again, we interpreted this result as being due the differences in the FWHM of the lines in regions where the velocity dispersion is higher, highlighting the spatial location where the gas is more turbulent.

\begin{figure*}
\resizebox{0.75\hsize}{!}{\includegraphics{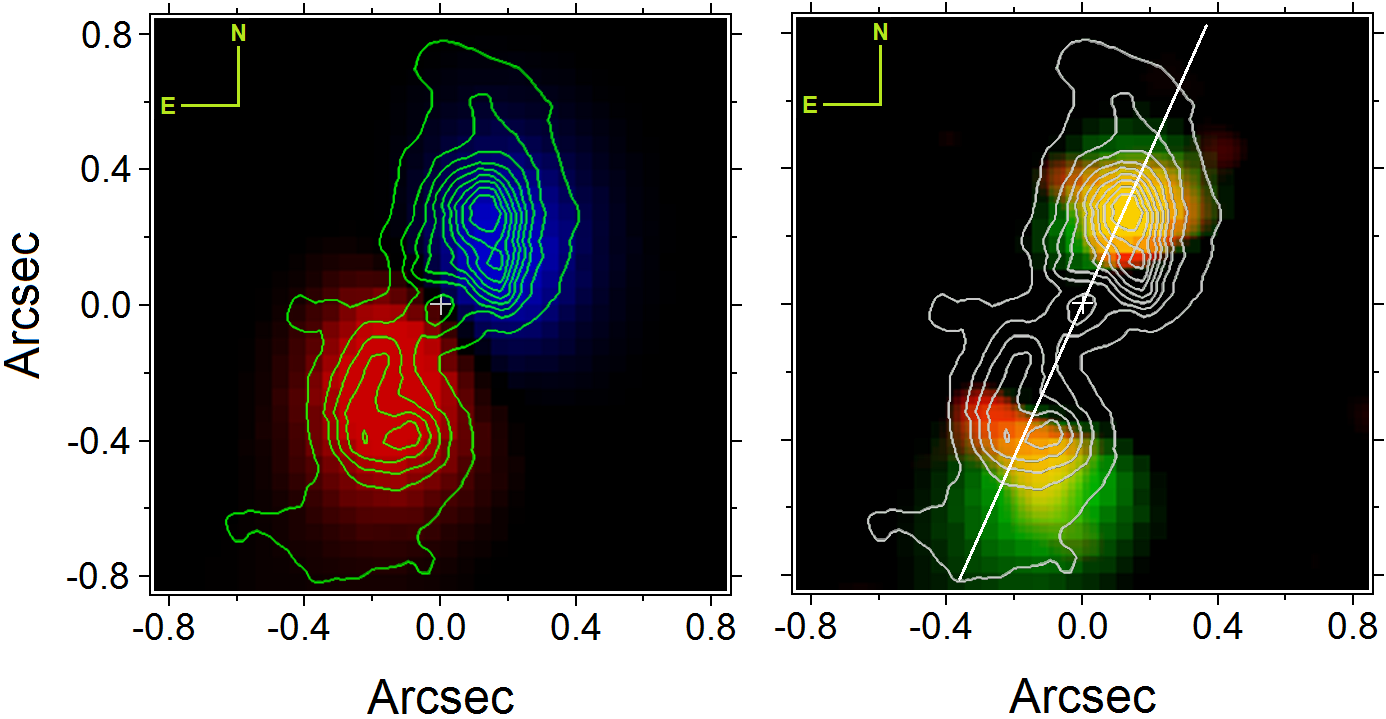}}
\caption{Left: tomogram 2 obtained with PCA tomography of the GMOS data cube, with the cones corresponding to positive and negative weight, shown in red and blue, respectively. Right: tomograms 3 for the GMOS (green) and NIFS (red); the yellow colour results from the mixing of the green and red colours. The white line indicates PA$_{Radio}$=156\textdegree. The contours of the (H$\alpha$+[N II])/I image are superposed on both images}
\label{fig:cone}
\end{figure*}

Fig.~\ref{fig:cone} (left) shows tomogram 2, spatially coincident with the image of ionized gas from the HST, although with a resolution $\sim$~6 times lower. This tomogram informs which side is approaching and which is moving away from us. 

It is important to emphasize that tomograms 2 and 3 reveal features that were clearly detected only by means of this technique. In Fig.~\ref{fig:cone} (right), we compare the third tomogram of PCA tomography from the NIFS and GMOS data cubes, which shows the corresponding regions of higher velocity dispersion. One can see they have similar morphologies and that the observed structures are located close to the edges of the ionization cones, along the position angle of the radio jet. Although observed with a significant difference in resolution, it is surprising that both turbulent components of ionized and molecular gas are located in the same region.

\subsection{The optical line ratios and the physical conditions}
\label{sec:lr}

\begin{figure}
 \begin{minipage}{0.95\columnwidth}
  \resizebox{\hsize}{!}{\includegraphics{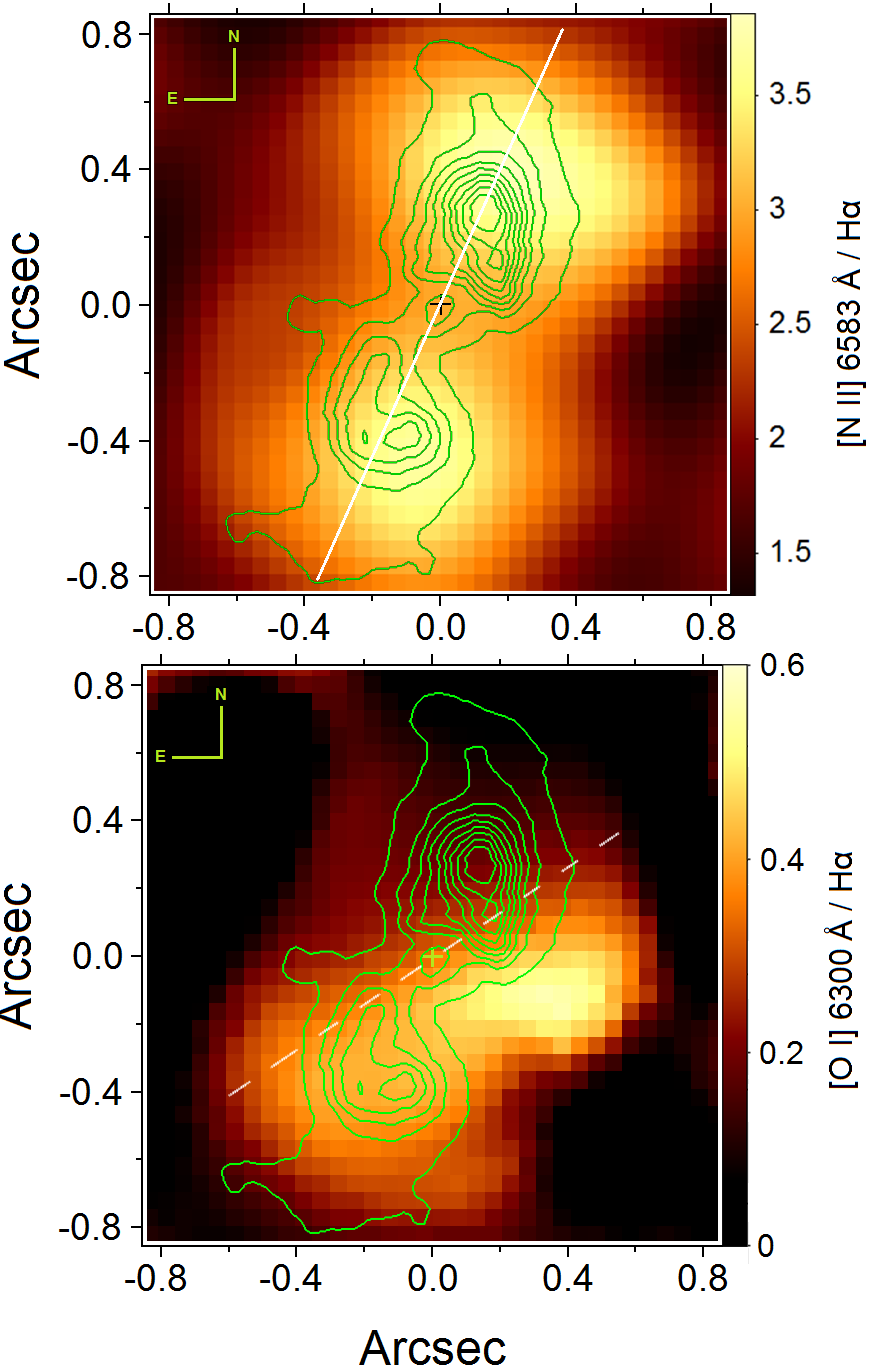}}
 \end{minipage}
 \caption{Top: image of the [N II]/H$\alpha$ ratio. The white line gives PA$_{radio}$=156\textdegree. Bottom: image of the [O I]/H$\alpha$. The dashed white line denotes the position angle (124\textdegree) of the molecular disc. Both images show the contours of the HST image.}
\label{fig:NIIHa}
\end{figure}

Other line ratios useful to investigate the excitation mechanism are [N II]$\lambda$6583 \AA/H$\alpha$ and [O I] $\lambda$6300 \AA/H$\alpha$ (Fig.~\ref{fig:NIIHa}), which can be calculated for each spaxel in the entire FOV, comprising the nuclear region and the outflow. The line profiles were fitted with a Gauss-Newton algorithm for non-linear functions. Fig.~\ref{fig:NIIHa} (top panel) shows the [N II]/H$\alpha$ ratio, together with the contours of the (H$\alpha$+[N II])/I image. Both images depict a bipolar structure, centred on the AGN, with similar position angles. But it is important to notice that the maximum [N II]/H$\alpha$ ratio does not coincide with the maximum emission of the ionization cones, represented by the HST contours. The ratios are in the range of [N II]/H$\alpha \sim$~1-3.8 and have a typical value of $\sim$2.5, with a maximum of $\sim$3.8 at 53 pc in the NW and SE directions from the nucleus. The same ratio was found for M51 ([N II]/H$\alpha\sim$3.8), which has a strong jet-cloud interaction \citep{Kuno96}. The line connecting the two regions with the highest values of [N II]/H$\alpha$ has PA$\sim$160\textdegree$\pm6$\textdegree, consistent with the PA of the radio emission. In contrast, the regions of star formation in the stellar ring have values between [N II]/H$\alpha=0.3-0.4$~\citep{Thaisa07}, typical of H II emission \citep{Baldwin81}.

The line ratio of [O I]/H$\alpha$ is shown in Fig.~\ref{fig:NIIHa} (bottom panel), together with the ionization cones, represented by the HST image and the position angle of the molecular disc. It is interesting to note the maximum ratio, 0.6, which does not coincide with the cones or the H$_{2}$ disc. This feature is not related to differences in extinction. In fact, it is located in a region where extinction should be higher (see Fig.~\ref{fig:hst}) and, therefore, with a lower [O I] emission. The peak of the spot is located 50 pc south-west from the nucleus, and its intensity decreases in the direction of the southern cone, right below the location of the molecular disc. However, there is some correlation between the high ratio of [O I]/H$\alpha$ with the far side of the ionization cone, in the south-east, but none in the another cone. The average ratio in the FOV (200 pc$^{2}$) is [O I]/H$\alpha$=0.24, in agreement with \citet{Ho97315}, who obtained a value of 0.23.

High values of [O I]/H$\alpha$ could be associated with mechanical heating by shocks \citep{Osterbrock06}, although are probably not related to the outflow for this galaxy, since the radio emission has a distinct orientation. If this result is, indeed, indicative of shock waves, there is no strong evidence of any young stellar population in the nucleus \citep{VanderLaan13}, which would be related with supernova remnants and stellar winds. A similar structure, of comparable dimension, was found near the nucleus in M81 \citep{Ricci15} and, on a smaller scale, in stellar clusters at the centre of the Milky Way, as the Arches Cluster \citep{Yusef02}, and IRS 16 \citep{Genzel10}, likely associated with stellar winds of young stars.

The reason we did not analyze the [S II]$\lambda$6717/$\lambda$6731 ratio is that these lines are absent in nearly half of the FOV.  

\subsection{The kinematics of the H$\alpha$ and [N II] lines}
\label{sec:ionizedkin}

\begin{figure*}
\resizebox{0.90\hsize}{!}{\includegraphics{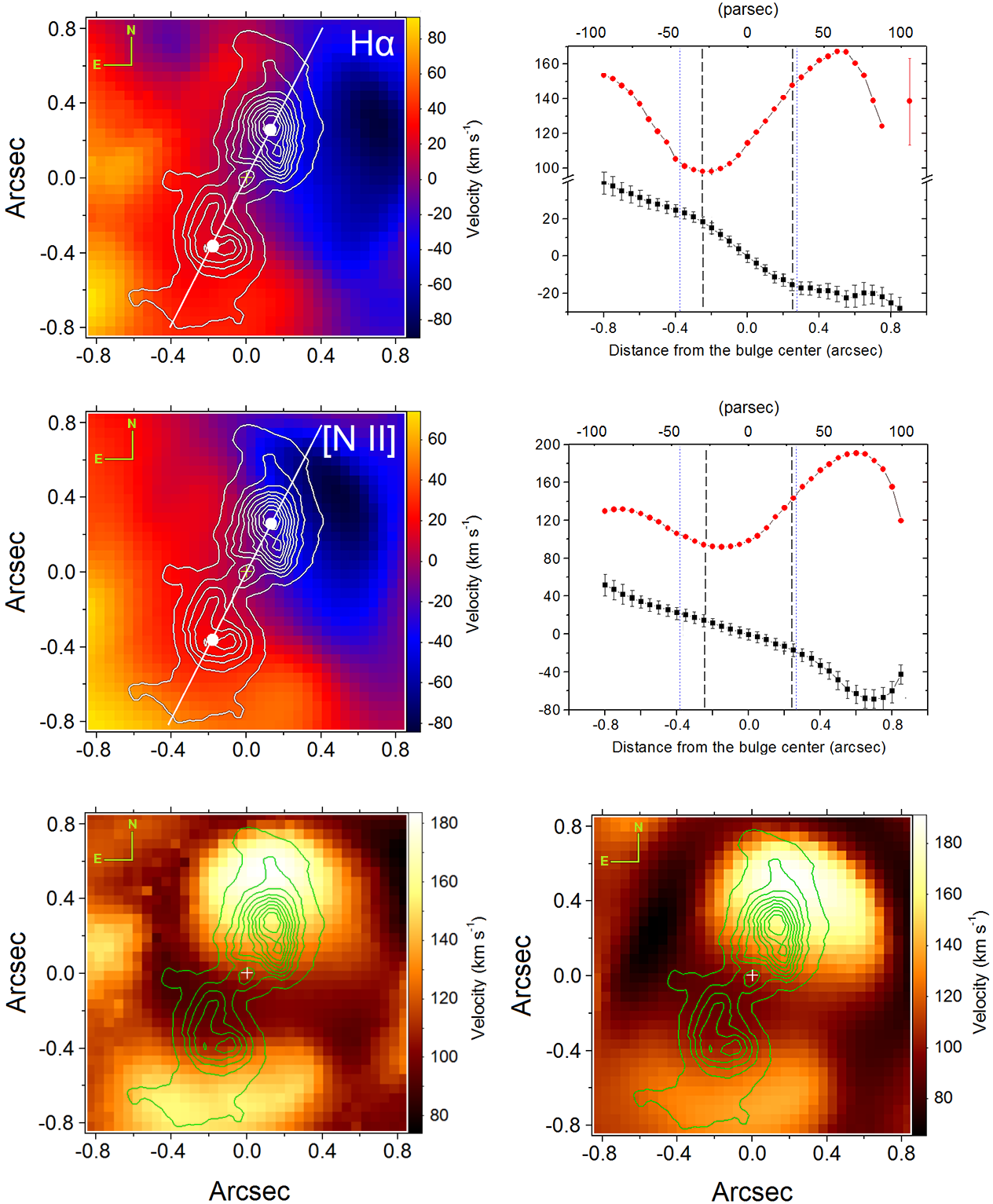}}
\caption{Top left: H$\alpha$ velocity map. Superposed on this image are the contours of the (H$\alpha$+[N II])/I image. The filled circles correspond to the regions of the bi-cone with highest intensity. The white line indicates PA$_{cone}$=154\textdegree $\pm1$\textdegree. Top right: radial velocity and velocity dispersion profiles (black squares and red circles, respectively) along the same PA. The error bar for the velocity dispersion, corrected for instrumental broadening, is shown within the graph. Middle panel: the same for the [N II] line. The vertical dashed black lines on both graphs denote the FWHM of the PSF and the dotted lines denote the location of the regions with higher intensity. Bottom left: H$\alpha$ velocity dispersion map. Bottom right: velocity dispersion map of the [N II] line.}
 \label{fig:NIIHakin}
\end{figure*}

The radial velocity and velocity dispersion maps for the H$\alpha$ and [N II] lines, with the instrumental broadening of $\sim$50 km s$^{-1}$, were obtained fitting a Gaussian function to the emission line profiles at each spaxel in the data cube. We show the results in Fig.~\ref{fig:NIIHakin}. The H$\alpha$ velocity ranges from 85 km s$^{-1}$ to -95 km s$^{-1}$ along the kinematic PA=121\textdegree$\pm$2\textdegree. For the same data, but with a FOV of $7^{\prime\prime}\times15^{\prime\prime}$, \citet{Thaisa07} found an upper limit of 220 km s$^{-1}$, and a PA=125\textdegree$\pm$10\textdegree, and \citet{Dumas07} measured a PA$_{[OIII]}$=142\textdegree$\pm$1\textdegree, within $33^{\prime\prime}\times41^{\prime\prime}$. The [N II] kinematics has consistent values of 75 km s$^{-1}$ to -85 km s$^{-1}$, with PA=103\textdegree$\pm$2\textdegree. The position angles were calculated by the method described in Appendix C of \citet{Krajinovic06}, using an \textsc{idl} program implemented by Michele Cappellari.
 
The velocity dispersion maps show, for both lines, a double peak symmetrically located along the same direction of the ionization cones, reaching a maximum of $\sim$184 km s$^{-1}$. \citet{Thaisa07} obtained velocity dispersions with typical values in the range of 60 to 80 km s$^{-1}$ for the [N II] and H$\alpha$ lines, reaching up to 140 km s$^{-1}$ in the ionization cones. The FOV presented here is smaller, and therefore the average velocity dispersion obtained was 111 km s$^{-1}$. The peak for the [N II] line is displaced from the H$\alpha$ peak by 26 pc and -7 pc along the $x$ and $y$ directions, respectively, and is closer to the PA of the radio emission. Neither of the peaks coincides with the peaks of the H$\alpha$+[N II] emission seen in the HST image but are, instead, located right after its extremities (which can be seen in the velocity dispersion measurements along the cones' PA, in the graphs of Fig.~\ref{fig:NIIHakin}). At the edge of the maps, the S/N is insufficient to draw any conclusions.

\begin{figure}
\resizebox{0.95\hsize}{!}{\includegraphics{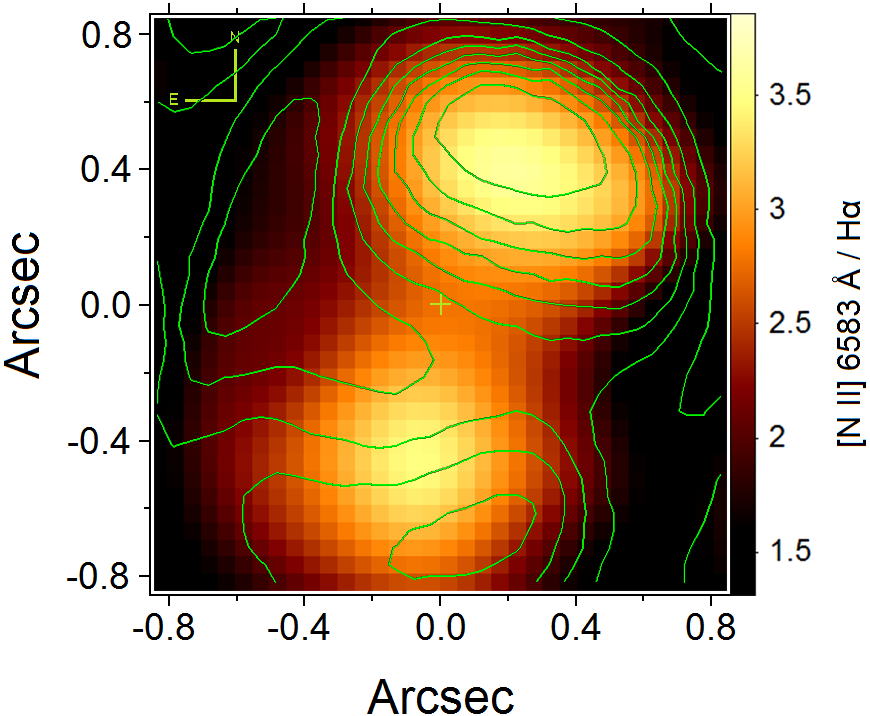}}
\caption{[N II]/H$\alpha$ ratio and the contours of the velocity dispersion map of the [N II] line.}
\label{fig:NII}
\end{figure}

The distribution for the [N II]/H$\alpha$ ratio, shown in Fig.~\ref{fig:NIIHa}, also depicts a bipolar structure, similar to that of the velocity dispersion (Fig.~\ref{fig:NIIHakin} bottom panel). In Fig.~\ref{fig:NII}, we can see a strong correlation between their structures, more so in the NW than in the SE direction. 

\section{The stellar component}
\label{sec:stellar}

\subsection{The Sersic index}
\label{sersic}

The Sersic index is a good discriminant between classical and pseudo-bulges. Most of the pseudo-bulges have $n<2$; classical bulges, $n>2$  (\citealt{Fisher08, Kormendy04}). In order to investigate if the galaxy bulge, in the central 200 pc, can be classified as a classical bulge or a pseudo-bulge, we fitted a 2D Sersic profile both to the NIFS K-band image and to the I-band image of the HST. We obtained indexes of 1.4 $\pm$0.1, and 2.1 $\pm$0.1 in the K-band and I-band, respectively. This region is much smaller than the effective radius for this galaxy ($R_{e}=$49$^{\prime\prime}$ in the B-band, \citealt{Marquez93}). This result, together with the fact that NGC 6951 hosts a bar connected to a prominent stellar ring and has spiral dust structures in the nucleus, lead to the conclusion that this galaxy has, indeed, a pseudo-bulge. 

Despite the fact that the bar in NGC 6951 connects to a nuclear ring of star formation, no young stellar population was found in the circumnuclear region between the nucleus and the ring, only a bulge-like profile of old stars, with age $>3$ Gyr \citep{VanderLaan13}. In addition to this old component, the optical images from the HST reveal spiral structures of gas that are not forming new stars. In order to check if any stellar substructure might appear, we subtracted the image of the bulge from the image of the exponential model with the indexes $n=1.8$ and $n=2.1$, in the K and I bands, respectively, but no stellar substructure was identified.

\subsection{The stellar kinematics}

\begin{figure*}
\resizebox{0.95\hsize}{!}{\includegraphics{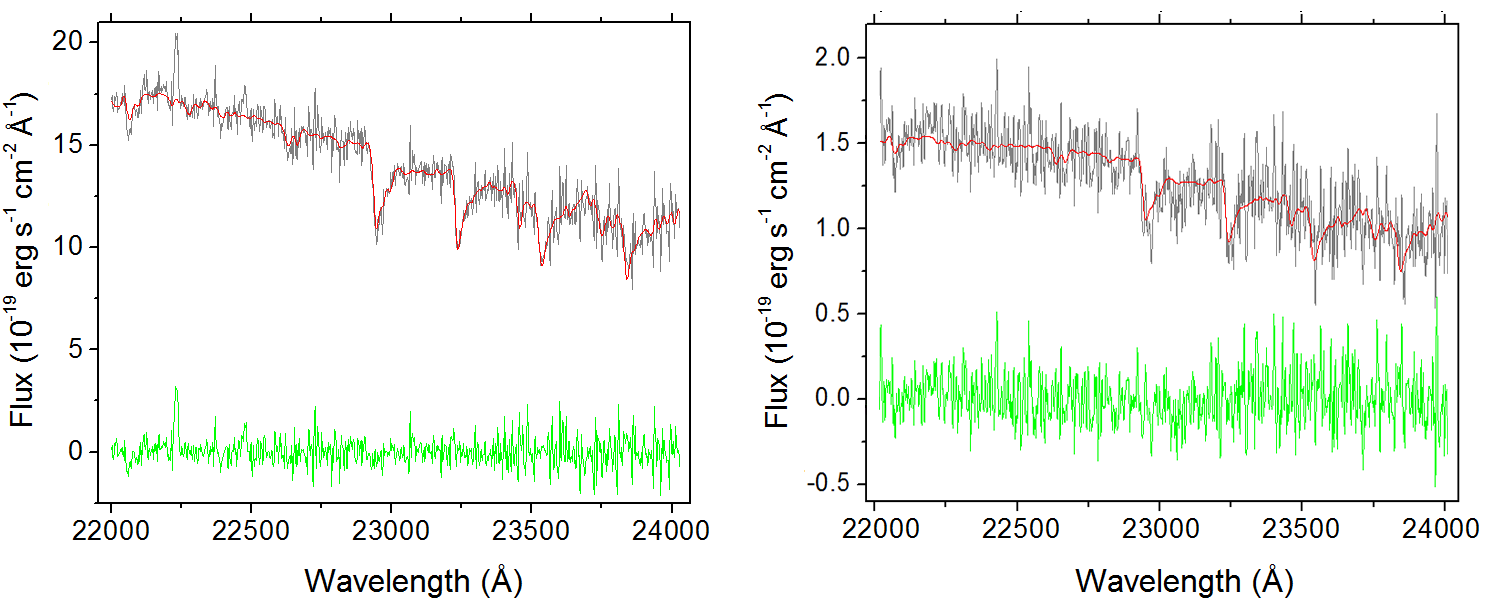}}
\caption{Left: nuclear spectrum of the central spaxel (black), with S/N=23 and the resulting fit using pPXF (red). The residual spectrum is shown in green. Right: the same as the left panel for one spaxel with S/N=10.}
\label{fig:stellarfit}
\end{figure*}

\begin{figure}
\begin{minipage}{0.43\textwidth}
  \resizebox{\hsize}{!}{\includegraphics{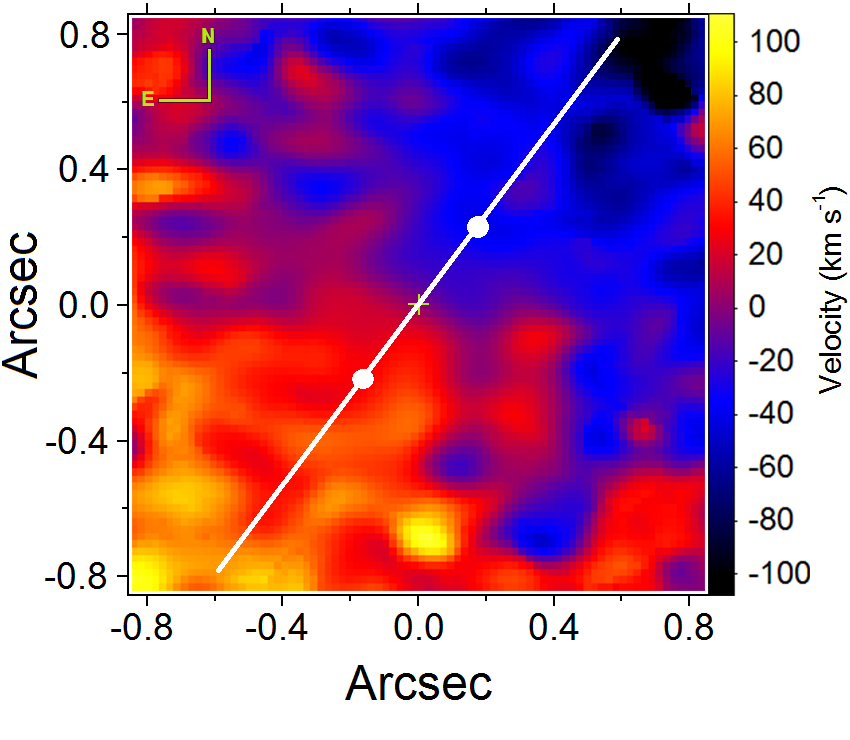}}
 \end{minipage}
 \begin{minipage}{0.43\textwidth}
  \resizebox{\hsize}{!}{\includegraphics{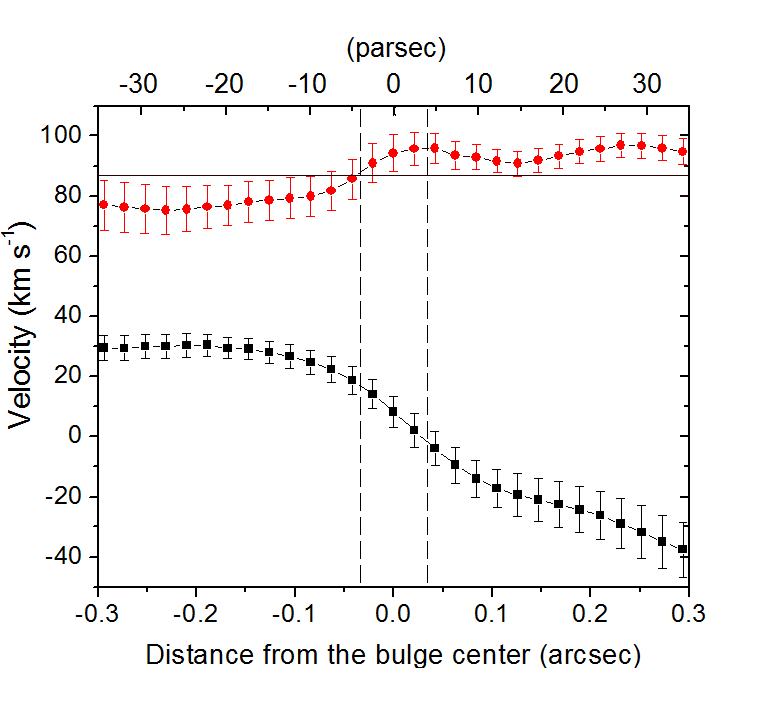}}
 \end{minipage}
\begin{minipage}{0.43\textwidth}
  \resizebox{\hsize}{!}{\includegraphics{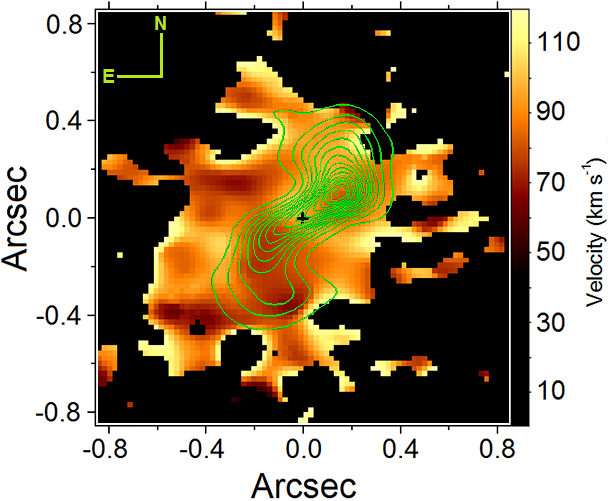}}
 \end{minipage}
 \caption{Top: stellar velocity map obtained from the pPXF fit. The corresponding kinematic axis has PA=143\textdegree~and the white filled circles in the image denote the range of the plot. Middle: radial velocity profile (black squares) and velocity dispersion (red squares) along the kinematic axis. The vertical dashed lines denote the FWHM of the PSF. Bottom: velocity dispersion map, corrected for instrumental broadening, with the contours of the molecular gas. The regions with S/N$\leq$10 are masked in black.}
\label{fig:stellarkin}
\end{figure}

To extract the stellar line-of-sight velocity distribution (LOSVD), we used the $^{12}$CO and $^{13}$CO stellar absorption bands to fit the best combination of a stellar spectra template \citep{Winge09}, convolved with a Gauss-Hermite series. This was done for each spaxel in the data cube after we had masked all the emission lines before the fitting procedure. The best-fitting parameters were computed by the \textsc{Penalized Pixel Fitting} (pPXF) method, implemented by \citet{Cappellari04}; at the end of the process each individual spaxel had a corresponding value for the velocity. In Fig.~\ref{fig:stellarfit}, we show the nuclear spectral fit of one single spaxel with S/N=23 (left panel), and a fit with S/N=10 (right panel). The average S/N in the nuclear region is $\sim$20 for an aperture radius of $\sim0^{\prime\prime}.1$, with a minimum of $\sim$7 at the borders of the data cube. In Fig.~\ref{fig:stellarkin} (top panel), one can see the radial velocity map for the stellar field, with kinematic PA=143\textdegree $\pm$2\textdegree, which agrees with the PA extracted from the low-resolution mode of the SAURON data \citep{Dumas07}, 145\textdegree $\pm$2\textdegree, for a FOV $\sim$20 times larger.

We measured a heliocentric radial velocity of $V_{r}$=1454$\pm$5 km s$^{-1}$, which corresponds to a redshift of $z=0.00485$. The projected velocity within $\sim0^{\prime\prime}.3$ is shown in Fig.~\ref{fig:stellarkin} (top panel), and ranges from -35 to +35 km s$^{-1}$. 
The middle panel of Fig.~\ref{fig:stellarkin} shows the velocity profile with a pseudo-slit of $\sim0^{\prime\prime}.1$ along the position angle of 143\textdegree, as well as the velocity dispersion for the same orientation. Along the pseudo-slit, a mean velocity dispersion of 87$\pm$8 km s$^{-1}$ was found. In addition, we detected a curious feature: velocity dispersion is lower and nearly constant for redshifted velocities (79$\pm$4 km s$^{-1}$), but rises $15\pm6$~km s$^{-1}$ for blueshifted velocities (94$\pm$2 km s$^{-1}$), where the near side of the outflow is located.

With regard to velocity dispersion (Fig.~\ref{fig:stellarkin}, bottom panel), we found an average of $97\pm3$ km s$^{-1}$, within $\sim0^{\prime\prime}.4$, weighted by intensity. This procedure ensures that the errors in the velocity dispersion for the pixels with low S/N have a lower weight. The pixels with S/N$\leq$10 were excluded from this calculus. The velocity dispersions obtained for spaxels corresponding to spectra with S/N$\leq$10 were greater than 120 km s$^{-1}$. This is an artifact of fitting under low S/N conditions. Therefore, these spaxels were masked in the velocity dispersion map. 

The ratio between the velocity dispersion of the ionized gas and of the stars in galaxies ranges from 0.6 to 1.4, with an average value of 0.80 \citep{Ho09699}, but here the dynamics of the ionized gas is clearly affected by the outflow, with an average velocity dispersion of $\sim$130 km s$^{-1}$ for the H$\alpha$ line, corrected for instrumental broadening. In fact, in the case of NGC 6951, \citet{Ho09699} adopted a stellar velocity dispersion estimated from the [N II] $\lambda$6583 line, overestimating its value, since this line is probably enhanced by the outflow dynamics.
The variation between the stellar and the H$\alpha$ PAs in the FOV is 24$\pm$4\textdegree, which is 17\textdegree~above the average difference observed in nine active galaxies \citep{Dumas07}. These larger misalignments are correlated with BHs accretion rates higher than $10^{-4.5}\Msun$ yr$^{-1}$.

\section{Discussion}
\label{sec:discussion}

Two important large-scale phenomena associated with AGNs, the inflow and outflow of gas, are supposed to be mostly present at the same time (\citealt{Martini99, Thaisa10, Davies14}), even if both are not always observable or well discriminated. In fact, the AGN feedback can be easily traced by the emission lines from the ionized gas, while evidences of inflowing gas are more difficult to detect, and the inflow rate is more difficult to measure. The inflow is necessary to feed the SMBH and accounts for the observed energy that is generated and transferred to the surrounding gas, which in turn is heated and blown away in the form of winds and/or jets. 

For NGC 6951, there is a claim of evidence of inflow seen in the H$\alpha$ kinematics \citep{Thaisa07}.  Evidence for outflow is seen in the ionization cones in the HST image as well as in the ionized gas emission of the NIFS and GMOS data cubes.

\subsection{The molecular disc as inflated by the jet}

There is a strong indication that NGC 6951 hosts a weak nuclear jet (with spectral index of $\sim$0.6, \citealt{Saikia02}), and no signal of a significant on-going star formation in the nucleus, which could imply in a radio emission from supernova remnants (\citealt{Perez00,VanderLaan13}). This emission has an angular size of $\sim0^{\prime\prime}.7\times0^{\prime\prime}.2$, at a position angle of 156\textdegree. Such dimensions comprise the region of the molecular emission ($\sim0^{\prime\prime}.8\times0^{\prime\prime}.6$) and the orientation is similar to that one of the detected turbulent regions (Fig.~\ref{fig:cone}, right panel).

We propose that the structure we see at the centre of NGC 6951 can be explained by two ionization cones, co-aligned with a radio jet. We also see a molecular disc that seems to be impacted by the jet; this collision ejects gas from the disc, heating it and increasing the turbulence. This gas, being blown into the cones, is ionized by the central source. In our interpretation, the jet is probably interacting with the inner edge of the rotating molecular disc.

\begin{figure}
\resizebox{0.99\hsize}{!}{\includegraphics{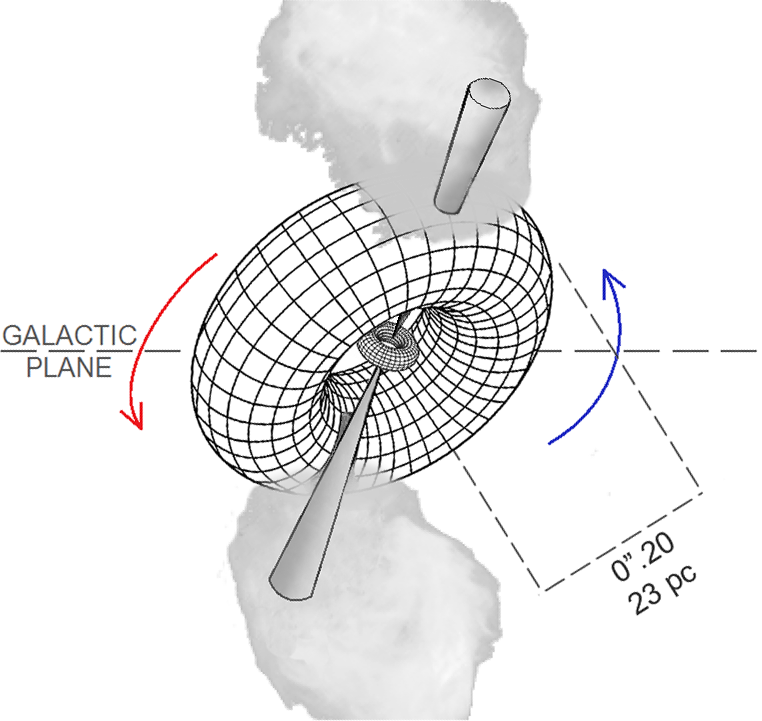}}
\caption{A sketch of the interaction between the jet and the molecular disc, inclined with respect to our LOS. The hypothetical torus, shown as the central disc, is aligned with the jet. The arrows denote the direction of the disc's rotation.}
 \label{fig:model}
\end{figure}

In the present case, we can safely claim only that the jet is inclined with respect the molecular disc; inclined enough to interact with its inner walls. Given our observations, it is not possible to predict how the gas takes its way from the detected disc down to the centre. Perhaps, the most natural interpretation is the existence of a dusty molecular torus (demanded by the Unified Model), which defines the orientation of the ionization cones and of the inner accretion disc. This interpretation requires the hypothetical torus to not be aligned with the observed molecular disc and, furthermore, that it should be edge-on, as we see no evidence of broad line emission.

There is no reason to stress the connection between the molecular disc and the supposed torus, since there is no indication of how or if they are physically connected. As pointed out in Section~\ref{sec:intro}, the discussion in \citep{Lawrence07} could be adapted to our scenario, where cloud instabilities, located in the molecular disc, give rise to episodes of inflow with random orientations. We are not excluding other possible complex configurations, as for example, the presence of warps in scales of the molecular disc \citep{Krips07} or of the accretion disc (\citealt{Greenhill03, Hernstein05, Kuo11}). However, a torus aligned with the jet is the most natural interpretation to explain the entire picture in light of the Unified Model.

Fig.~\ref{fig:model} compares this model with our actual data: the direction of the jet given by the hypothetical inclined inner accretion disc interacting with the inner walls of the molecular disc. It is worth mentioning that both ionization cones, as seen in the HST and GMOS data, and the regions of greater gas turbulence are aligned with the jet (Figs.~\ref{fig:6regionsprofiles1}, ~\ref{fig:h2pca}, ~\ref{fig:hsth2}, ~\ref{fig:cone} and ~\ref{fig:NIIHa}). The kinematics of the cones was revealed by PCA tomography (Fig.~\ref{fig:cone}) as well as by velocity measurments of H$\alpha$ and [N II] (Fig.~\ref{fig:NIIHakin}), with the side of approach corresponding to the NW region from the nucleus, and therefore to the near side of the cone/jet.

The turbulent gas associated with the high velocity dispersion regions likely consists of partially ionized gas, with the ionization provided by the central source. X-ray emission from the AGN may also play a role in heating the gas \citep{Halpern83}. In addition, heating may also be enhanced either by high-velocity electrons from the jet or by shocks from the turbulence. This may be the reason why the [N II]/H$\alpha$ ratio is higher at this position (Fig.~\ref{fig:NII}).
 
Since the gas excavated from the molecular disc is blown into the ionization cones, there is an interesting question to discuss: what is the fraction of the gas in the cones that originates from the molecular disc and how much is blown out from the central part of the AGN? We do not see how to answer this question based on our observations alone. 

Complementing this scenario, \citet{Elvis10} have shown that primarily feedback mechanisms, as radiation pressure from the central source, could generate instabilities and dramatically increase the cross-section of cold clouds, which strengthen the gas outflow. On the other hand, the density and the cloud fragmentation in the disc is probably high enough to stop the jet (\citealt{Gaibler11,Wagner12}). In this sense, NGC 6951 may be a downscaled version of NGC 1068, where a similar interaction can be found. In the latter case, a large-scale jet is bended by molecular clouds \citep{Gallimore96}. The location where the jet is deflected is heated and sweeps away the gas in form of bubbles and small accelerating clumps, possibly driven by gas pressure as a secondary outflow (May D. \& Steiner J. E., in preparation).

In NGC 6951, both the molecular disc and the hypothetical dusty torus are inclined, nearly edge-on (although not co-aligned). This shows that neither of them is aligned with the disc of the galaxy itself, inclined by 46\textdegree. 
Misalignments between outflows and discs of galaxies were found to be common in more than 50\% of a sample of active galaxies \citep{Fischer13}, and maybe in four out of five active galaxies studied by \citep{Davies14}, implying that the orientation of the torus has no correlation with the galactic disc.

\subsection{The molecular mass outflow rate}

We can estimate the mass of hot H$_{2}$ gas in the outflow, represented by regions 3 and 4 in Fig.~\ref{fig:h2flux} (left panel), which in principle is the real amount of gas being ejected, without any fraction of cold gas. Following the calculations performed by \citet{Scoville82} and \citet{Riffel08}:

\begin{equation}
\begin{aligned}
M_{H_{2}}&=\frac{2m_{p}F_{H_{2}\lambda 2.1218}4\pi D^{2}}{f_{\nu =1,J=3}A_{S(1)}h\nu}\nonumber \\
&=5.0776\times 10^{13}~\left(\frac{F_{H_{2}\lambda 2.1218}}{erg s^{-1}cm^{-2}}\right)\left(\frac{D}{Mpc}\right)^{2}
\label{h2mass}
\end{aligned}
\end{equation}

where $m_{p}$ is the proton mass, $F_{H_{2}\lambda 2.1218}$ is the line flux (not corrected for extinction and assumed to have a small effect; see Sect~\ref{sec:hst}), $D$ is the galaxy distance and $f_{\nu =1,J=3}$ is the fraction of hot H$_{2}$ in the level $\nu=1$ and $J=3$ , with $M_{H_{2}}$ given in solar masses. The linear dependence of the H$_{2}$ emissivity on density derives from the assumption of a thermalized gas, at 2000 K and with $n_{H_{2}}>10^{4.5}$ cm$^{-3}$. This implies a population fraction of $1.22\times10^{-2}$ with transition probability $A_{S(1)}=3.47\times10^{-7}$ s$^{-1}$. Using Table~\ref{table:flux}, for the NW outflow region 4, we have $F_{H_{2}\lambda 2.1218}=4.24\pm0.12 \times 10^{-16} erg~s^{-1} cm^{-2}$ within an aperture radius of $\sim0^{\prime\prime}.1$, so we obtain $M_{H_{2}}\sim12\Msun$. Performing the same calculation in region 3, we again obtain $M_{H_{2}}\sim12\Msun$, resulting in $\sim 24\Msun$ of hot gas removed from the molecular disc. By calculating the hot H$_{2}$ mass in the disc, corresponding to regions 1 and 2, we obtain a total of $M_{H_{2}}\sim37\Msun$. Considering the total mass of hot molecular gas, 40\% of the gas being emitted is in the outflow.

On the other hand, we can also estimate the mass outflow rate by taking the characteristic outflow speed as $V\sim\sqrt{v^{2}+\sigma^{2}}/\langle\sin\rangle$, with $\langle\sin\rangle=0.7$, since we do not know the orientation of the outflowing H$_{2}$ (see \citealt{Davies14}), obtaining $V\sim135$ km s$^{-1}$. Assuming the average distance of 25 pc from the AGN and the total mass of the hot molecular gas of $24\Msun$ for the outflow, we obtain a molecular mass outflow rate of $\sim10^{-4} \Msun~yr^{-1}$. Given the high misalignment between the disc and the outflow, if we assume that all the outflowed gas is coming from the disc and is not part of the ISM, we can set this value as the minimum inflow rate required to maintain the molecular structure, in terms of the hot H$_{2}$ mass. This fact leads to the interesting conclusion that if the mass inflow rate is smaller than this, the molecular disc will eventually disappear.

\subsection{The dynamics of the central molecular gas}
\label{sec:dyn}

In order to analyze if there is some connection between the molecular gas in the stellar ring and the region delimited by the ring, \citet{VanderLaan11} fitted a second Gaussian component to the CO(2-1) emission in the flux map, in addition to that of the bar model, and found what they called a ``CO bridge'', interpreted as gas inflow through the disc, with a PA of $\sim$23\textdegree~(see Fig.7, bottom left, of \citealt{VanderLaan11}). The detection is the only hint, so far, of molecular gas inflowing inside the stellar ring with a defined velocity. This CO component does not reach the centre, but is connected to the detected HCN central emission, which has a similar velocity range of $\pm$70 km s$^{-1}$ and PA of 160\textdegree $\pm 20$\textdegree~(Fig.2 of \citealt{Krips07}). This PA is significantly different from that of the major axis of the galaxy (PA$=135$\textdegree). The non-alignment may be caused by a different inclination of the central gas disc, by non-circular velocities or even by a warp, since the kinematics for the HCN and the CO bridge seem to behave as if they are in the disc.

The HCN flux map has a resolution $\sim10$~times lower when compared to our data. It is also a compact structure with nearly the same extension of the H$_{2}$ emission, but remains unresolved in this case. Its kinematic axis has a PA closer to that of the radio emission than of the H$_{2}$ disc, indicating that the extended HCN gas is probably not related to the detected H$_{2}$ disc. \citet{Krips07} claimed that this could be the a circumnuclear disc/torus, a hypothesis that is not incompatible with our results, but we were able to see with better resolution an edge-on H$_{2}$ disc with an upper limit for the scale height of about 20 pc.

The emission outside the detected molecular structure is too low, preventing a reliable link between the extended H$_{2}$ kinematics at the border of our FOV and the CO and HCN kinematics. However, it is clear that the kinematics of the warm H$_{2}$ disc is uncoupled from the mapped CO and the central HCN gas, based on the inclination of the H$_{2}$ disc with respect to the galaxy disc of $\sim44$\textdegree~(considering $i_{H_{2}}\sim90$\textdegree~and $i_{Gal}$=46\textdegree). \citet{Krips07}, based on their mass estimates, argued that the HCN emission comes from a disc with low inclination. 
Even if the potential of the ring can drive gas to the centre, the H$_{2}$ emission reveals a different history about how this inflow takes place in the central 50 pc, since the disc inclinations are considerably different. There is no change in the kinematic position angle of the nuclear disc to consider the presence of a warped disc. In order to explain the apparent lack of correlation between the nuclear disc and the larger scale galactic disc, \citet{Hopkins12} carried out a set of high-resolution simulations of the circumnuclear region. In the absence of a secondary bar structure, they claim that large-scale fragmentation of the gas in the galactic disc can lead to misaligned nuclear discs; however, their simulations show discs with smaller dimensions than the one we found.

\subsection{The black hole mass}
\label{sec:bh}

Since the H$_{2}$ disc has a very steep inclination, assumed to be 90\textdegree, and a radial velocity curve typical of discs (Fig.~\ref{fig:h2vel}), we can estimate an upper limit for the BH mass within the radius where the velocity curve peaks. The dynamic effect of the velocity dispersion is not included in the calculation because the regions where we find an increase in dispersion are at a different position angle, and we had assumed that the molecular gas in the disc, which is in thermal equilibrium, presents only local turbulence. Taking the radius of $\sim0^{\prime\prime}.15$, which corresponds to $\sim17$ pc in the galaxy, and a velocity of 40 km s$^{-1}$, we obtain a dynamic mass of $6.3\times10^{6}$\Msun. This value is about 1.5 times the measured mass of the Milky Way ($\sim4.4\times10^{6}$\Msun; see \citealt{Genzel10}), without subtracting the stellar mass within this radius. The total mass of gas was found to be less than 3\% of the stellar mass in the nuclei of a sample of 6 galaxies in \citet{Mazzalay13}.

Based on dynamical modeling of the gas with the central ($\sim0^{\prime\prime}.2$) emission-line width measured with the HST, taking $i=33$\textdegree~and a stellar velocity dispersion of 104 km s$^{-1}$, \citet{Beifiori09} found an upper limit of $5.9\times10^{6}$\Msun. Their result comprises a radius slightly larger, but gives a mass 7\% lower than the one we estimated. The dynamical mass estimated by \citet{Krips07}, in the central $\sim0^{\prime\prime}.5$~of the HCN emission, was $2\times10^{8}$\Msun~for an inclination of 40\textdegree, which is more than one order of magnitude above our result, for a radius $\sim3\times$~larger. We should remember that the HCN kinematic position angle is very similar to the radio position angle, which in our data accounts only for the kinematics of the outflow. Even if the HCN is in a disc with inclination similar to the galaxy, it is possible that both circular motions and the outflow kinematics are superposed at their measured velocities. This would lead to an overestimation of the dynamical mass. 

As pointed out by \citet{Kormendy13}, galaxies with pseudo-bulges, such as NGC 6951, do not follow the $M_{BH}-\sigma$~relation (Sect.~\ref{sersic}). Pseudo-bulges have smaller BH mass at a given $\sigma$~than galaxies with classical bulges. Nevertheless, as a matter of completeness, applying the $M_{BH}-\sigma$~relation of \citet{Ferrarese00}, with the central stellar velocity dispersion of 97 km s$^{-1}$, in an aperture radius of $\sim0^{\prime\prime}.2$, gives us $M_{BH}=4.3\times10^{6}$\Msun.
This value is 47\% lower than the one we obtained, in contradiction with pseudo-bulges having lower masses than what is provided by the $M_{BH}-\sigma$~relation. This discrepancy comes mainly from the fact that our estimate still implies in a significant amount of stellar mass.

\section{Conclusions}
\label{sec:conclusions}

We have presented and analyzed the high-resolution NIR data cube of NGC 6951 obtained with the NIFS spectrograph on the Gemini North Telescope. We also re-analyzed archive data both from the HST (images) and the Gemini North GMOS (data cubes) of the same galaxy. We focused on the central 200 pc. Our results are based both on the molecular and on the ionized gas phases, and the main conclusions are the following: 
\\
\begin{enumerate}
\item We detected a compact structure of molecular gas seen in H$_{2}$, interpreted as a nearly edge-on disc with diameter of $\sim$47 pc, PA=124\textdegree~and velocity range from -40 to +40 km s$^{-1}$.  This disc, probably the source of the gas that feeds the AGN, is misaligned with respect to the radio jet emission, which has a PA=156\textdegree. 
\\
\item The position angle of the radio jet is consistent with that of the ionization cones, with PA=153\textdegree$\pm$2\textdegree, seen in the H$\alpha$+[NII] HST image. The H$\alpha$+[NII] emission is also seen in the GMOS data cube; the brightest side is blueshifted, revealing that ionized structures are seen in the outflow.
\\
\item There are two regions of turbulent gas, seen both in the molecular and the ionized phases, that are connected to the edges of the molecular disc. These two turbulent spots have a PA similar to that of the radio jet and of the ionization cones.
\\
\item The two turbulent spots coincide with high [NII]/H$\alpha$ ratios of $\sim$3.8, suggesting that the region is excited by shocks.
\\
\item Our results, which show the correlation between the velocity dispersion of the [N II] line and the [N II]/H$\alpha$ line ratio, agree with the work of \citet{Thaisa07}, indicating that the [N II] excitation is related to its kinematics, although this correlation is better seen in the NW part of the cone.
\\
\item Based on the H$_{2}$ line ratios, we conclude that the excitation mechanism is mainly due to shocks, with a temperature of 1980$\pm$130 K, compatible with thermal equilibrium of the molecular gas.
\\
\item We explain the molecular structure as being one thick, rotating disc connected to two turbulent regions, derived from a ``digging process'' that the jet inflicts on the internal parts of the disc, ejecting some of the molecular gas. The molecules are exposed to the ionizing cones, and they probably dissociate to form ionized gas in the cones. As a consequence, the excavated molecular and ionized gas phases are both turbulent. This seems to be the most evident feedback activity in this object.
\\
\item A self-consistent interpretation of our data requires the existence of an edge-on dusty torus, as demanded by the Unified Model, that hides the BLR. This torus defines the orientation of the ionization cones and is inclined with respect the molecular disc.
\\
\item The HST (V-I) image shows an irregular distribution of dust, not related to the molecular emission. This suggests that most of the gas may be too cold to be detected.
\\
\item The velocity curve along the molecular disc suggests that the dynamical mass within 17 pc is $M_{dyn}=6.3\times10^{6}$\Msun, establishing an upper limit to the central BH mass.
\end{enumerate}

These findings were only possible because of the combination of high-resolution data and accurate image processing techniques. Together, these observations set up a new consistent scenario for the inner 200 pc dynamics of NGC 6951. The molecular lines of H$_{2}$ are so far the best indicators of gas inflow, because the molecules start to emit in the region of the most relevant physical process inherent to the AGN, while the kinematics for the ionized gas is mostly described by the outflow feedback. 

\section*{Acknowledgments}

The authors are grateful for the insightful suggestions of
the anonymous referee, which have improved this manuscript. This work is based on observations obtained at the Gemini Observatory, which is operated by the Association of Universities for Research in Astronomy, Inc., under a cooperation agreement with the NSF on behalf of the Gemini partnership: the National Science Foundation (United States), the Science and Technology Facilities Council (United Kingdom), the National Research Council (Canada), CONICYT (Chile), the Australian Research Council (Australia), Minist\'{e}rio da Ci\^{e}ncia, Tecnologia e Inova\c{c}\~{a}o (Brazil) and CONICET (Argentina). This work is also based on observations made with the NASA/ESA \textsc{Hubble Space Telescope} obtained at the Space Telescope Institute, which is operated by the Association of Universities for Research in Astronomy, Inc., under NASA contract NAS5-26555. Finally, we would like to thank FAPESP for support under grants 2011/19824-8 (DMN), 2011/51680-6 (JES), 2008/06988-0 (TVR), 2012/02268-8 (RBM) and 2011/20223-9 (ISA).

\bibliographystyle{mnras}
\bibliography{ngc1068} 




\label{lastpage}

\end{document}